\documentclass[preprint,showpacs,preprintnumbers,amsmath,amssymb]{revtex4}
\usepackage[abs]{overpic}
\usepackage{graphicx,subfigure,mathrsfs}
\begin{document}
\preprint{HUPD1305}
\def\tbr{\textcolor{red}}
\def\tcr{\textcolor{red}}
\def\ov{\overline}
\def\dprime{{\prime \prime}}
\def\nn{\nonumber}
\def\f{\frac}
\def\H{\mathcal{H}}
\def\beq{\begin{equation}}
\def\eeq{\end{equation}}
\def\bea{\begin{eqnarray}}
\def\eea{\end{eqnarray}}
\def\bsub{\begin{subequations}}
\def\esub{\end{subequations}}
\def\dc{\stackrel{\leftrightarrow}{\partial}}
\def\ynu{y_{\nu}}
\def\ydu{y_{\triangle}}
\def\ynut{{y_{\nu}}^T}
\def\ynuv{y_{\nu}\frac{v}{\sqrt{2}}}
\def\ynuvt{{\ynut}\frac{v}{\sqrt{2}}}
\def\d{\partial}
\def\sla#1{\rlap/#1}
\def\mH{\mathscr{H}}
\def\tD{\tilde{D}}
\def\Q{{\cal Q}}
\title{Quark sector CP violation of the universal seesaw model}
\author{Ryomu Kawasaki}
\author{Takuya  Morozumi}
\email[E-mail: ]{morozumi@hiroshima-u.ac.jp}
\author{Hiroyuki Umeeda}
\email[E-mail: ]{umeeda@theo.phys.sci.hiroshima-u.ac.jp}
\affiliation{Graduate School of Science,
Hiroshima University, Higashi-Hiroshima, 739-8526, Japan}
\def\nn{\nonumber}
\def\beq{\begin{equation}}
\def\eeq{\end{equation}}
\def\bei{\begin{itemize}}
\def\eei{\end{itemize}}
\def\bea{\begin{eqnarray}}
\def\eea{\end{eqnarray}}
\def\ynu{y_{\nu}}
\def\ydu{y_{\triangle}}
\def\ynut{{y_{\nu}}^T}
\def\ynuv{y_{\nu}\frac{v}{\sqrt{2}}}
\def\ynuvt{{\ynut}\frac{v}{\sqrt{2}}}
\def\s{\partial \hspace{-.47em}/}
\def\ad{\overleftrightarrow{\partial}}
\def\ss{s \hspace{-.47em}/}
\def\pp{p \hspace{-.47em}/}
\def\bos{\boldsymbol}
\begin{abstract}
We study the charge parity (CP) violation of the universal seesaw model, especially its quark sector. The model is based on $SU(2)_L \times SU(2)_R \times U(1)_{Y^\prime}$. 
In order to count the number of parameters in the quark sector, we use the degree of freedom of the weak basis transformation. For the $N(3)$-generation model, the number of 
CP violating phases in the quark sector is identified as $3N^2-3N+1$ $(19)$. 
We also construct 19 CP violating weak basis invariants of Yukawa coupling matrices and $SU(2)$ singlet quark
mass matrices in the three-generation universal seesaw model.
The quark interaction terms induced by neutral currents are given as 
an exact formula.
Both the charged current and the neutral current are expressed in terms of the mass basis by finding the transformations from the weak basis to the mass basis.
Finally, we calculate the mixing matrix element approximately,
assuming that the $SU(2)_R$ breaking scale $v_R$ is much larger than
the electroweak breaking scale $v_L$.
\end{abstract}
\maketitle
\def\U{{\cal U}}
\def\D{{\cal D}}
\def\V{{\cal V}}
\def\Z{{\cal Z}}
\section{Introduction}
The universal seesaw mechanism [1-7]
based on $SU(3)_C\times SU(2)_R\times SU(2)_L\times U(1)_{Y^\prime}$
gauge symmetry is considered for fermion mass hierarchy with $SU(2)_R\times SU(2)_L$
isosinglet fermion masses.
The ordinary fermion and the singlet fermion mix at the tree level after spontaneous symmetry breaking
$SU(3)_C\times SU(2)_R\times SU(2)_L\times U(1)_{Y^\prime}\rightarrow SU(3)_C\times U(1)_{EM}$.
The universal seesaw mechanism provides us a clue for the mystery:
why are ordinary fermions much lighter than electroweak scale except for top quark [8,9]?
When this mechanism works, all of the strength of the Yukawa couplings can be taken 
order of unity.
The doublet quark and singlet quark are transformed by $SU(3)_C\times SU(2)_R\times SU(2)_L\times U(1)_{Y^\prime}$
as follows:
\bea
q_L\sim(3,1,2,\frac{1}{6}),\quad q_R\sim(3,2,1,\frac{1}{6}),
\quad \U\sim(3,1,1,\frac{2}{3}),\quad \D\sim(3,1,1,-\frac{1}{3})\nn
\eea
where $Q=T_R^3+T_L^3+Y^\prime$.
\\
\quad
A sophisticated discussion of CP violation using weak basis (WB) invariants
is given by Jarlskog in Ref.\cite{Jarlskog:1985ht} and by Bernabeu $et\: al$. in 
Ref.\cite{Bernabeu:1986fc}.
See also Ref.\cite{CPviolation} for a review and Ref.\cite{Branco:1986pr}
for WB invariants  in the framework of the left-right symmetric model.
 
The gauge boson mass matrix in the universal seesaw model is identical to
the left-right symmetric model studied in Ref. \cite{Chay:1998hd}, 
except that the left-right symmetric model includes the $SU(2)_R\times SU(2)_L$ bidoublet Higgs.
One can find the gauge boson mass matrix in the present model
by taking the limit where the vacuum expectation value of bidoublet Higgs vanishes.
The possibility that the universal seesaw mechanism resolves the strong CP problem is explained by
Babu and Mohapatra in Ref.\cite{Babu:1989rb}.
Embedding the universal seesaw in the grand unified theory scenario is discussed by Cho in Ref.\cite{Cho:1993jb}
, Koide in Ref. \cite{Koide:1998km}, and Mohapatra in Ref.\cite{Mohapatra:1996iy}.
\\
\quad
In this paper, we focus on the CP violation of the quark sector.
Phenomenological aspects of the CP violation have been studied in 
Refs. [19,20].
In the literature \cite{Kiyo:1998zm}, CP violation of the present model
is studied with an additional assumption: left-right symmetry.
We study the CP violation and the flavor mixing
as general as possible so that one can study the phenomenology
of the present model to the full extent.
The recent study on mixings of the vectorlike quarks can be also 
found in Ref. \cite{Aguilar-Saavedra:2013qpa}.

Our paper is organized as follows.
We count the number of the parameters in the quark sector in Sec. II.
In Sec. III, we construct WB invariants of the quark sector. 
In Sec. IV,
we propose a parametrization for the three-generation model
by minimizing the numbers of the parameters with weak basis transformation
(WBT). The relation between the WB invariant
and CP violation parameters in the specific parametrization is 
discussed.
The exact formulas for
the mixing matrices are obtained in the mass basis in Sec. V.
Finally, in Sec. VI, we carry out the diagonalization of 
$6 \times 6$ mass matrices with some approximation
and write down the mixing matrix elements.
Section VII is devoted to the summary.
\section{Counting the number of real and imaginary parameters in the quark sector of the universal seesaw model}
In this section, by using the freedom of WBT,
we minimize the number of real and imaginary parts of Yukawa couplings and singlet quark mass matrices.
The number of imaginary parts which are left after WBT corresponds to the number of physical CP violating phases. 
We also verify the number of 
CP violating phases by counting the independent number of 
CP invariant conditions in a specific weak basis.
\subsection{WBT of the universal seesaw model}
We assume the singlet 
quark generation number is $N$, which is
identical to an ordinary quark generation number.
In this model, WBTs on singlet and doublet quarks are given by
\begin{eqnarray}
&{\cal U}_{R}^{\prime}=V_{U_R}{\cal U}_{R},\quad
{\cal U}_{L}^{\prime}=V_{U_L}{\cal U}_{L},&\\
&{\cal D}_{R}^{\prime}=V_{D_R}{\cal D}_{R},\quad
{\cal D}_{L}^{\prime}=V_{D_L}{\cal D}_{L},&\\
&q_{R}^{\prime}=V_{R}q_{R},\quad
q_{L}^{\prime}=V_{L}q_{L},&
\end{eqnarray}
where ${\cal U}_{R(L)}$, ${\cal D}_{R(L)}$, and $q_{R(L)}$ denote 
the right-handed (left-handed) uptype singlet quark, downtype singlet quark, and ordinary doublet quark, respectively.
Below, the matrices with superscript $^\prime$ imply 
the matrices obtained by changing the WB.
Yukawa matrices and mass matrices of the singlet quarks are transformed as
\begin{eqnarray}
M_{\cal U}^{\prime}=V_{U_L}^{\dagger}M_{\cal U}V_{U_R},\nn \\
M_{\cal D}^{\prime}=V_{D_L}^{\dagger}M_{\cal D}V_{D_R},\nn \\
y_{uL}^{\prime}=V_L^{\dagger}y_{uL}V_{U_R},\nn \\
y_{uR}^{\prime}=V_R^{\dagger}y_{uR}V_{U_L},\nn \\
y_{dL}^{\prime}=V_L^{\dagger}y_{dL}V_{D_R},\nn \\
y_{dR}^{\prime}=V_R^{\dagger}y_{dR}V_{D_L},
\label{wbt}
\end{eqnarray}
where $M_{{\cal U}({\cal D})}$ denotes the $N \times N$ 
uptype (downtype) mass matrix of a singlet quark, and
$y$ is the $N \times N$ Yukawa coupling constant matrix.
One chooses the weak basis, and $M_{{\cal U}({\cal D})}^\prime$ is given by
a real diagonal matrix by carrying out the suitable biunitary transformation
as the WBT. In the basis,
both of uptype and downtype singlet mass matrices have $N$ real parameters.
Suppose that we find the biunitary transformation, which diagonalizes 
the mass matrices as
\begin{eqnarray}
\tilde{V}_{U_L}^{\dagger}M_{\cal U}\tilde{V}_{U_R}=D_U,\\
\tilde{V}_{D_L}^{\dagger}M_{\cal D}\tilde{V}_{D_R}=D_D,
\label{eq:Diagonal}
\end{eqnarray}
where $D_U$ and $D_D$ are real diagonal matrices.
We note that real diagonal matrices are invariant under the 
similarity transformation $P_U$ and $P_D$,
\bea
P_U^\dagger D_U P_U=D_U, \quad P_D^\dagger D_D P_D=D_D,
\eea
where $P_U$ and $P_D$ are given by,
\bea
P_U=
\left(
\begin{array}{cccc}
\mathrm{e}^{ia_1}& & & \\
&\mathrm{e}^{ia_2} & & \\
& &\ddots & \\
& & &\mathrm{e}^{ia_N}
\end{array}
\right),\quad
P_D=
\left(
\begin{array}{cccc}
\mathrm{e}^{ib_1}& & & \\
&\mathrm{e}^{ib_2} & & \\
& &\ddots & \\
& & &\mathrm{e}^{ib_N}
\end{array}
\right).
\eea
Unitary matrices which diagonalize the singlet quark mass matrices with biunitary transformation are not fixed uniquely. 
One can define the new unitary matrices, 
\bea
V_{U_R}=\tilde{V}_{U_R} P_U, \quad V_{U_L}=\tilde{V}_{U_L} P_U, \nn \\ 
V_{D_R}=\tilde{V}_{D_R} P_D, \quad V_{D_L}=\tilde{V}_{D_L} P_D.
\eea   
By using  $V_{U_R},V_{U_L}$, $V_{D_R}$, and $V_{D_L}$ as WBT, one can also
diagonalize the singlet quark mass matrices.
Next we consider the weak basis transformation on Yukawa matrices,
\begin{eqnarray}
&&V_L^{\dagger}y_{uL}V_{U_R}\; = 
\; P_U^{\dagger}(\tilde{V}_L^{\dagger}y_{uL}\tilde{V}_{U_R})P_U,
\label{eq:yul} \\
&& V_R^{\dagger}y_{uR}V_{U_L}\; = 
\; P_U^{\dagger}(\tilde{V}_R^{\dagger}y_{uR}\tilde{V}_{U_L})P_U,
\label{eq:yur} \\
&& V_L^{\dagger}y_{dL}V_{D_R}\; = 
\; P_U^{\dagger}(\tilde{V}_L^{\dagger}y_{dL}\tilde{V}_{D_R})P_D,
\label{eq:ydl} \\
&& V_R^{\dagger}y_{dR}V_{D_L}\; = 
\; P_U^{\dagger}(\tilde{V}_R^{\dagger}y_{dR}\tilde{V}_{D_L})P_D.
\label{eq:ydr}
\end{eqnarray}
In Eqs.(\ref{eq:yul})-(\ref{eq:ydr}),  
we extract the diagonal phase matrix $P_U$
from  $V_L$ and $V_R$,
\bea
V_L=\tilde{V}_L P_U, \quad V_R=\tilde{V}_R P_U.
\eea
We can choose unitary matrix  $\tilde{V}_L$, so that 
$y^\prime_{\Delta_{uL}}=\tilde{V}_L^\dagger y_{uL}\tilde{V}_{U_R}$ is a 
lower triangular matrix with 
real diagonal elements.
One can also choose $\tilde{V}_R$, so that 
$y^\prime_{\Delta_{uR}}=
\tilde{V}_R^\dagger y_{uR}\tilde{V}_{U_L}$
is a lower triangular matrix with real diagonal elements.
Therefore, Eqs.(\ref{eq:yul}) and (\ref{eq:yur}) are rewritten as 
\begin{eqnarray}
&V_L^{\dagger}y_{uL}V_{U_R}=P_U^{\dagger}y^\prime_{\Delta_{uL}}P_U=
y_{\Delta_{uL}}, \nn \\
&V_R^{\dagger}y_{uR}V_{U_L}=P_U^{\dagger}y^\prime_{\Delta_{uR}}P_U=
y_{\Delta_{uR}}.
\label{wbt2} 
\end{eqnarray}
In the triangular form of the Yukawa couplings 
$y^\prime_{\Delta_{u L(R)}}$,  one reduces
$\displaystyle\frac{1}{2}N(N-1)$ real parameters and 
$\displaystyle\frac{1}{2}N(N+1)$ imaginary parameters 
from $N \times N$ complex Yukawa matrices $y_{uL}$ and $y_{uR}$, respectively.
Therefore, each triangular matrix includes $\displaystyle\frac{1}{2}N(N+1)$ real parts and 
$\displaystyle\frac{1}{2}N(N-1)$ imaginary parts. 
With $P_U$, one can remove the $N-1$ imaginary parts in $y^\prime_{\Delta_{uL}}$. Therefore, with the WBT in Eq.(\ref{wbt2}), $y_{\Delta_{uL}}$ includes
$ \displaystyle\frac{1}{2}N(N+1)$ real parts and 
$\displaystyle\frac{1}{2}(N-1)(N-2)$ imaginary parts, while  $y_{\Delta_{uR}}$ includes
$ \displaystyle\frac{1}{2}N(N+1)$ real parts and 
$\displaystyle\frac{1}{2}N(N-1)$ imaginary parts.

Next we count the number of parameters in $y_{dL}$ and $y_{dR}$.
We can use the similarity transformation $P_D$. Then one removes $N$ imaginary
parts in $y_{dL}$. Therefore, $y_{dL}$ includes 
$N^2$ real parts and $N^2-N$ imaginary parts.
Since we have already used all the freedom of WBT,
$N^2$ real parts and $N^2$ imaginary parts are
left in $y_{dR}$.

We summarize the number of degrees of freedom in the quark sector of
the universal seesaw model for $N$ generations.
Table \ref{tab1} shows the number of real and imaginary parameters in
the matrices obtained by the WBT.
Table \ref{tab2} shows the number of real and imaginary parameters for
specific generation numbers $N=1-4$. 
\subsection{CP invariant condition}
Let us prove the previous derivation of the number of 
CP violating phases with an alternative argument.
To count the numbers of nontrivial CP violating phases,  
one can study the numbers of independent CP invariant conditions.
The CP invariant conditions are then
\begin{eqnarray}
&&M_{\cal U}^{\prime}=M_{\cal U}^*,\quad
M_{\cal D}^{\prime}=M_{\cal D}^*,\\
&&y_{uL}^{\prime}=y_{uL}^*,\quad
y_{uR}^{\prime}=y_{uR}^*,\\
&&y_{dL}^{\prime}=y_{dL}^*,\quad
y_{dR}^{\prime}=y_{dR}^*.
\end{eqnarray}
We consider these conditions in a specific weak basis.  In the basis,
the singlet quark mass matrices 
are given by real diagonal matrices $D_{U}$ and $D_{D}$.
Yukawa coupling matrices
$y_{uL}$ and $y_{uR}$ are given by the
lower triangular matrices $y^\prime_{\Delta uL}$ and $y^\prime_{\Delta uR}$.
Note that the diagonal elements of the triangular matrix are real.
In this basis,  CP invariant conditions
for singlet quark mass matrices are written as
\bea
V_{U_L}^{\dagger}D_{U}V_{U_R}=D_{U},\quad
V_{D_L}^{\dagger}D_{D}V_{D_R}=D_{D}.
\eea
To satisfy the conditions given above, $V$s are determined as
\bea
V_{U_L}=V_{U_R}=P_U,\quad V_{D_L}=V_{D_R}=P_D.
\eea
The CP invariant conditions for Yukawa matrices are then 
\begin{eqnarray}
P_U^{\dagger}{y^\prime}_{\Delta_{uL}}P_U={y^\prime}_{\Delta_{uL}}^*,\\
P_U^{\dagger}{y^\prime}_{\Delta_{uR}}P_U={y^\prime}_{\Delta_{uR}}^*,\\
P_U^{\dagger}y_{dL}P_D=y_{dL}^*,\\
P_U^{\dagger}y_{dR}P_D=y_{dR}^*.
\end{eqnarray}
These four relations are also written in terms of the argument of their matrix element,
\begin{eqnarray}
\mathrm{arg}({y^\prime}_{\Delta_{uL}ij})&=&
\mathrm{arg}({y^\prime}_{\Delta_{uR}ij})= 
\frac{a_i-a_j}{2},
\label{eq:ytu} \\
\mathrm{arg}(y_{{dL}ij})&=& \mathrm{arg}(y_{{dR}ij})= \frac{a_i-b_j}{2}.
\label{eq:ytd}
\end{eqnarray}
We count the nontrivial CP invariant conditions which cannot be
satisfied by adjusting the phases in $P_U$ and $P_D$.
Since one can choose the $N-1$ phase difference, 
$\displaystyle{a_i-a_1}$ $(i=1-N)$ as 
$\mathrm{arg}(y_{\Delta uL\: i1})=\displaystyle\frac{a_i-a_1}{2}$,
the $N-1$ CP invariant conditions are automatically 
satisfied. Therefore, the number of the nontrivial conditions in
 Eq.(\ref{eq:ytu}) is $(N-1)^2=2 \times
\frac{N(N-1)}{2}-(N-1)$. 
As for the conditions in Eq.(\ref{eq:ytd}), $b_i$ is chosen
as $b_i=a_i-2\mathrm{arg}(y_{dLii})$ so that the $N$ condition of 
Eq.(\ref{eq:ytd}) is satisfied. Therefore, there are $2 N^2-N$
nontrivial conditions.  Then, in total, we find $3 N^2-3 N +1$
CP invariant conditions, which are identical to the number
of CP violating phases.
It also agrees with the number of the imaginary parts
in the Yukawa matrices obtained with the WBT (see Table \ref{tab1}).

\begin{table}
\caption{The number of parameters included in quark sector matrices for the $N$ generations universal seesaw model in a specific WB.}
\begin{tabular}{|c|c|c|c|c|c|c|c|}
\hline
 & $M_U$ & $M_D$ & $y_{\Delta uL}$&$y_{\Delta uR}$&$y_{dL}$&$y_{dR}$& Sum. \\ \hline
Re.&N&N&$\frac{1}{2}N(N+1)$&$\frac{1}{2}N(N+1)$&$N^2$&$N^2$&$3N(N+1)$ \\ \hline
Im.&0&0&$\frac{1}{2}(N-1)(N-2)$&$\frac{1}{2}N(N-1)$&$N(N-1)$&$N^2$&$3N^2-3N+1$
\\
\hline
\end{tabular}
\label{tab1}
\end{table}
\begin{table}
\caption{The number of parameters for the specific generation number $N$.}
\begin{tabular}{|c|c|c|c|c|}
\hline
 & $N=1$ & $N=2$ & $N=3$&$N=4$\\ \hline
Re.&6&18&36&60\\ \hline
Im.&1&7&19&37\\
\hline
\end{tabular}
\label{tab2}
\end{table}
\section{CP violating weak basis invariants in the three-generation model}
In this section, we derive the CP violating WB invariants for a three-generation model.
The use of the WB invariants including $SU(2)$ singlet quarks
within the standard model gauge group is discussed in Ref. \cite{Branco:1986my}.
We define the following Hermitian matrices in order to write down the WB invariants for CP violation in universal seesaw model:
\bea
&H_U=M_\U M_\U^\dagger,\quad
H_D=M_\D M_\D^\dagger,\quad
H_{uL}=y_{uL}y_{uL}^\dagger,&\nn\\
&H_{uR}=y_{uR}y_{uR}^\dagger,\quad
H_{dL}=y_{dL}y_{dL}^\dagger,\quad
H_{dR}=y_{dR}y_{dR}^\dagger,&\nn\\
&h_U=M_\U^\dagger M_\U,\quad
h_D=M_\D^\dagger M_\D,\quad
h_{uL}=y_{uL}^\dagger y_{uL},&\nn\\
&h_{uR}=y_{uR}^\dagger y_{uR},\quad
h_{dL}=y_{dL}^\dagger y_{dL},\quad
h_{dR}=y_{dR}^\dagger y_{dR}.&
\eea
In the case that the singlet quark generation number is 3, identical to the ordinary quark generation number, the 19 CP violating WB invariants in the quark sector of the universal seesaw model are then 
\bea
I_1&=&\mathrm{Imtr}[h_{uL},h_U]^3,
\label{I1}\\
I_2&=&\mathrm{Imtr}(M_\U h_Uh_{uL}M_\U^\dagger h_{uR})
,\label{I2}\\
I_3&=&\mathrm{Imtr}(M_\U h_U^2h_{uL}M_\U^\dagger h_{uR})
,\label{I3}\\
I_4&=&\mathrm{Imtr}(M_\U h_U^2h_{uL}H_U M_\U^\dagger h_{uR})
,\label{I4}\\
I_5&=&\mathrm{Imtr}[h_{dL},h_D]^3
,\label{I5}\\
I_6&=&\mathrm{Imtr}(M_\D h_D h_{dL}M^\dagger_\D h_{dR})
,\label{I6}\\
I_7&=&\mathrm{Imtr}(M_\D h_D^2h_{dL}M^\dagger_\D h_{dR})
,\label{I7}\\
I_8&=&\mathrm{Imtr}(M_\D h_D^2h_{dL}H_D M^\dagger_\D h_{dR})
,\label{I8}\\
I_9&=&\mathrm{Imtr}[H_{uL},H_{dL}]^3
,\label{I9}\\
I_{10}&=&\mathrm{Imtr}[H_{uR},H_{dR}]^3
,\label{I10}\\
I_{11}&=&\mathrm{Imtr}(M_\U y_{uL}^\dagger y_{dL}M_\D ^\dagger y_{dR}^\dagger y_{uR})
,\label{I11}\\
I_{12}&=&\mathrm{Imtr}(M_\U y_{uL}^\dagger y_{dL}M_\D^\dagger H_D y_{dR}^\dagger y_{uR})
,\label{I12}\\
I_{13}&=&\mathrm{Imtr}(M_\U y_{uL}^\dagger y_{dL}M_\D^\dagger H_D^2y_{dR}^\dagger y_{uR})
,\label{I13}\\
I_{14}&=&\mathrm{Imtr}(M_\U h_Uy_{uL}^\dagger y_{dL}M_\D^\dagger y_{dR}^\dagger y_{uR})
,\label{I14}\\
I_{15}&=&\mathrm{Imtr}(M_\U h_Uy_{uL}^\dagger y_{dL}M_\D^\dagger H_D y_{dR}^\dagger y_{uR})
,\label{I15}\\
I_{16}&=&\mathrm{Imtr}(M_\U h_U y_{uL}^\dagger y_{dL}M_\D^\dagger H_D^2y_{dR}^\dagger y_{uR})
,\label{I16}\\
I_{17}&=&\mathrm{Imtr}(M_\U h_U y_{uL}^\dagger y_{dL}M_\D^\dagger y_{dR}^\dagger y_{uR})
,\label{I17}\\
I_{18}&=&\mathrm{Imtr}(M_\U h_U y_{uL}^\dagger y_{dL}M_\D^\dagger H_D y_{dR}^\dagger y_{uR})
,\label{I18}\\
I_{19}&=&\mathrm{Imtr}(M_\U h_Uy_{uL}^\dagger y_{dL}M_\D^\dagger H_D^2y_{dR}^\dagger y_{uR}).
\label{eq:CPVwb}
\eea
We briefly explain how to construct the CP violating WB invariants in Eqs.(\ref{I1})-(\ref{eq:CPVwb}).
First, we can construct the WB invariant which does not vanish trivially  by considering the trace of the cube of the commutator,
\beq
I_1=\mathrm{Im tr}[h_{uL},h_U]^3.
\eeq
Note that the real part of the trace of the cube of the commutator 
does vanish.  The nonzero
value of the trace of the
cubic commutator signals CP violation, and the proof follows in the same way
as the Jarlskog invariant \cite{Jarlskog:1985ht} and 
the CP violating WB invariant \cite{Bernabeu:1986fc}
for the Kobayashi-Maskawa
model \cite{Kobayashi:1973fv}.
Next we consider the WB invariant with the form,
\beq
\mathrm{tr}(M_\U h_Uh_{uL}M_\U^\dagger h_{uR}).
\label{WBinvariant}
\eeq
When CP is conserved, the imaginary part of Eq. (\ref{WBinvariant}) vanishes,
\beq
\mathrm{tr}(M_\U^*h_\U^*h_{uL}^*M_\U^T h_{uR}^*)=[\mathrm{tr}(M_\U h_Uh_{uL}M_\U^\dagger h_{uR})]^*.
\eeq
Therefore, the imaginary $I_2=\mathrm{Imtr}(M_\U h_Uh_{uL}M_\U^\dagger h_{uR})$ is a CP violating WB invariant.
By inserting some Hermitian matrices, we can also construct the other CP violating WB invariants.
\section{A parametrization of the Yukawa sector in the three-generation model}
In Sec. II, we introduced a specific WB; i.e., the uptype Yukawa matrices
are given by the triangular matrices and the singlet quark matrices are real
diagonal.
This WB is obtained by fully utilizing the freedom of the WBT.
Then the number of the real parts and imaginary parts included in the parameters of the Yukawa sector is minimized and should be equal to the number of independent physical parameters.
In this section, we introduce a parametrization of the Yukawa sector for the three-generation model which is associated with the WB
in Table I. 
The parametrization includes the same number of the
real and imaginary parameters with that of the WB for $N=3$.
The Yukawa terms for the quarks in the WB are given by the following Lagrangian:
\bea
{\cal L}_{\mathrm{Yukawa}}&=&y_{\Delta uLij}\overline{q^i_L} \tilde{\phi}_L \U^j_R
+y_{\Delta uRj i}^\ast \overline{\U^i_L} \tilde{\phi}_R^\dagger q^j_{R} 
+\overline{\U_L^i}
 D^i_{U} \U_R^i + h.c.\nn \\
&+&y_{dLij} \overline{q^i_L} \phi_L \D^j_R
+y_{dRj i}^{\ast} \overline{\D^i_L} \phi_R^\dagger q^j_{R} 
+\overline{\D_L^i}
 D^i_{D} \D_R^i+h.c.,
\eea 
where $i,j=1-3$.
After the symmetry breaking of $SU(2)_L$ and $SU(2)_R$, the doublet Higgses
$\phi_L$ and $\phi_R$ acquire the vacuum expectation values $v_L$ and $v_R$,
respectively. Then the mass matrix for six up (down) quarks are generated as
\bea
&& \begin{pmatrix} \overline{u_L} & \overline{\U_L} \end{pmatrix}
{\cal M_U} \begin{pmatrix} u_R \\ \U_R \end{pmatrix}, \quad
\begin{pmatrix} \overline{d^0_L} & \overline{\D_L} \end{pmatrix}
{\cal M_D}^0 \begin{pmatrix} d^0_R \\ \D_R \end{pmatrix},
\eea
where, 
${\cal M_U}$ and ${\cal M_D}^0$ are given as
\bea
{\cal M_U}&=&\begin{pmatrix} 0 & y_{\Delta uL} v_L\\ 
                        y_{\Delta uR}^\dagger v_R & D_U \end{pmatrix}
\label{eq:Bigu}, \\
{\cal M_D}^0&=&\begin{pmatrix} 0 & y_{dL} v_L\\ 
                        y^{\dagger}_{dR} v_R & D_D \end{pmatrix}.
\eea
$D_U$ and $D_D$ are singlet quark mass matrices
which are real diagonal,
\bea
D_U=\begin{pmatrix} M_U & 0 & 0 \\
                     0   & M_C & 0 \\
                     0   & 0 & M_T \end{pmatrix}
,\quad
D_D=\begin{pmatrix} M_D & 0 & 0 \\
                     0   & M_S & 0 \\
                     0   & 0 & M_B \end{pmatrix},
\eea
where the diagonal elements satisfy the following order,
$M_U > M_C > M_T$ and $M_D > M_S >M_B$,
in order to acquire the light quark mass spectrum $m_{u}<m_{c}<m_t$
and $m_d<m_s<m_b$.
Applying the result in Table \ref{tab1} to the three-generation model,
the uptype Yukawa matrices 
$y_{\Delta uL}$ and $y_{\Delta uR}$ are given as triangular matrices,
\bea
y_{\Delta uL}=\begin{pmatrix} y_{uL1} & 0 & 0 \\
                       y_{uL21} & y_{uL2} & 0 \\
                       y_{uL31} & y_{uL32} & y_{uL3} 
\end{pmatrix},
\quad
y_{\Delta uR}=\begin{pmatrix} y_{uR1} & 0 & 0 \\
                       y_{uR21} & y_{uR2} & 0 \\
                       y_{uR31} & y_{uR32} & y_{uR3} 
\end{pmatrix},
\label{eq:productup}
\eea
where $y_{uL32}$ and $y_{uRij}$($i > j$) are complex 
and the other elements are real. 
Two phases of $y_{uL21}$ and $y_{uL31}$
are removed by using the freedom of the similarity transformation
$P_U$ in Eq.(\ref{eq:yul}). 
The downtype Yukawa couplings are given by $3 \times 3$ matrices.
According to Table \ref{tab1}, $y_{dL}$ includes nine real parts
and six imaginary parts.  They can be parametrized as
\bea
y_{dL}= U_L y_{\Delta dL},
\label{eq:product}
\eea
where $y_{\Delta dL}$ is a lower triangular matrix [$
(y_{\Delta dL})_{ij}=0
$, for ($i<j$)] which 
includes six real parts and only one imaginary part in $
{(y_{\Delta dL})}_{32}$.
It is parametrized exactly the same as
that of $y_{\Delta uL}$,
\bea
y_{\Delta dL}=\begin{pmatrix} y_{dL1} & 0 & 0 \\
                       y_{dL21} & y_{dL2} & 0 \\
                       y_{dL31} & y_{dL32} & y_{dL3} 
\end{pmatrix}.
\eea
$U_L$ includes three angles and five phases as
\bea
U_L&=&P(\alpha_{L1}, \alpha_{L2}, 0) V(\theta_{L1}, \theta_{L2},\theta_{L3}, \delta_L) P(\beta_{L1}, \beta_{L2},0), 
\label{eq:UL}
\\
&& P(\phi_1,\phi_2,\phi_3)=\begin{pmatrix} e^{i \phi_1} & 0 & 0 \\
                                         0 & e^{i \phi_2} & 0 \\
                                         0 & 0 & e^{i \phi_3} 
\end{pmatrix}.
\label{eq:P}
\eea 
In $U_L$, $V$ denotes the Kobayashi-Maskawa-type parametrization
of the unitary matrix which includes three mixing angles $\theta_{Li}$
($i=1-3$) and a single
CP violating phase $\delta_L$. [see Eq.(\ref{eq:ckm}) in Appendix for the
explicit form for $V$]. 
There are four more CP violating phases, $\alpha_{Li},\beta_{Li}$ $(i=1,2)$,
which are parametrized in the diagonal phase matrix in $P(\alpha_{1L},\alpha_{2L},0)$ and $P(\beta_{L1}, \beta_{L2},0)$ in Eq.(\ref{eq:P}).
Next we parametrize the down-quark Yukawa coupling $y_{dR}$. Since $y_{dR}$ is a completely general $3 \times 3$ complex matrix,  it has three more CP violating phases compared with $y_{dL}$. 
Therefore one  can parametrize it as the product of a unitary matrix and triangular matrix as
\bea
y_{dR}=U_R y_{\Delta dR}. 
\label{eq:productR}
\eea
In the parametrization
given in Eq.(\ref{eq:productR}), the unitary matrix 
$U_R$ includes six phases [see Eq.(\ref{eq:param})],
\bea
U_R&=& P(\alpha_{1R}, \alpha_{2R}, \alpha_{3R})
V(\theta_{1R}, \theta_{2R}, \theta_{3R}, \delta_R)
P(\beta_{R1}, \beta_{R2}, 0). 
\eea
$y_{\Delta dR}$ has the same form as that of $y_{\Delta uR}$,
\bea
y_{\Delta dR}&=&\begin{pmatrix} y_{dR1} & 0 & 0 \\
                       y_{dR21} & y_{dR2} & 0 \\
                       y_{dR31} & y_{dR32} & y_{dR3} 
\end{pmatrix},
\eea
where $y_{dR ij} (i>j)$ are complex and $y_{dRi} (i=1,2,3)$ are real.

We show how the 19
CP violating WB invariants $ I_1-I_{19}$ 
in Eqs.(\ref{I1})-(\ref{eq:CPVwb})
can be written in
the specific WB in which the singlet quark mass matrices
are real diagonal and the Yukawa couplings are parametrized
by Eqs. (\ref{eq:productup}), (\ref{eq:product}),
and Eq.(\ref{eq:productR}).
Then one can relate the CP violating WB invariants to the CP violating 
parameters defined by the specific WB. We first show that the first eight WB invariants $I_1-I_8$ can be written in terms of 
the CP violating phases of the
Yukawa couplings of the triangular matrices. Note that there are also eight CP
violating phases in the triangular matrices of the Yukawa couplings. 
By taking the real diagonal mass matrices for
the singlet quarks, one can show $I_1$ is written in terms
of a combination of the Yukawa coupling $y_{\Delta uL}$, 
\bea 
I_1 \ni {\rm Im}\Bigl{[}h_{uL 12} h_{uL 23} h_{uL 31}\Bigr{]},
\eea
where
$h_{uL}=y_{\Delta uL}^\dagger y_{\Delta uL}$. 
Because ${\rm Im}{(y_{uL32})}$ is 
the only CP violating phase in $y_{\Delta uL}$, 
$I_1$ corresponds to the CP violating phase $ {\rm Im} (y_{uL32})$.
One can also show that $I_2, I_3,$ and $I_4$ are written by linear
combinations of the following quantities:
\bea
\chi_u^{ij}={\rm Im}{(h_{uL ij} h_{uR ji})}, (i,j)=(1,2),(2,3),(3,1), 
\eea
where  $h_{uR}=y_{\Delta uR}^\dagger y_{\Delta uR}$.
$I_2,I_3,$ and 
$I_4$ depend on  ${\rm Im} (y_{uR ij})$ $(i>j)$
and ${\rm Im} (y_{uL 32})$.  All the four CP violating phase in uptype
Yukawa couplings $y_{\Delta uL}$ and $y_{\Delta uR}$ can be found in 
the WB invariants $I_1-I_4$. Similarly, the four WB
invariants $I_5-I_8$ are related to the four CP violating phases
in the triangular matrices of the down quark sector. 
$I_5$ is related to ${\rm Im}(y_{dL32})$ 
since $I_5$ is proportional to
\bea
I_5  \ni {\rm Im}(h_{dL12}h_{dL23}h_{dL31}),
\eea   
where $h_{dL}=y_{\Delta d}^\dagger y_{\Delta d}$.
$I_6-I_8$ are written in terms of three combinations
of Yukawa couplings 
$\chi_{d}^{12}$,$\chi_{d}^{23}$, and $\chi_{d}^{31}$. They are defined by
\bea
\chi_d^{ij}={\rm Im}{(h_{dL ij} h_{dR ji})}, 
\quad (i,j)=(1,2),(2,3),(3,1). 
\eea
where $h_{dR ji}=y_{\Delta d R}^\dagger y_{\Delta d R}$.
They are related to ${\rm Im}(y_{\Delta dRij})$ ($i>j$) 
and ${\rm Im}(y_{dL 32})$.

So far, all the CP violating phases in the
triangular matrices in the Yukawa couplings are identified in 
the WB invariants $I_1-I_8$. 
Next, we show how the
other 11 WB invariants are related to the 
rest of the CP violating phases in $U_L$ and $U_R$.
Although $I_9-I_{19}$ depend on the CP
violating phases of the triangular matrices, we focus on 
their dependence on the CP violation of unitary matrices $U_L$
and $U_R$. $I_9$ depends on $U_L$ and $I_{10}$ depends on $U_R$.
There are still four CP violating phases in $U_L$ and five CP
violating phases which are not identified yet in the WB invariants.
One can easily see $I_{11}-I_{19}$ can be written in terms of 
\bea
{\rm Im} (y_{\Delta uL}^\dagger U_L y_{\Delta dR})_{ij} 
         (y_{\Delta dR}^\dagger U_R^\dagger y_{\Delta uR})_{ji}. 
\eea
They depend on the 11 CP violating phases in $U_L$ and $U_R$.

Now we carry out the following unitary transformations on 
the downtype quarks
$d^0_L$ and $d^0_R$,
\bea
d_L^0&=&U_L d_L,
\label{eq:downL}\\
d_R^0&=&U_R d_R.
\label{eq:newbasis}
\eea
With the new basis given in  
Eqs.(\ref{eq:downL}) and (\ref{eq:newbasis}), only the form of 
charged currents changes as
\bea
&&\overline{u_L} \gamma_\mu d^0_L=\overline{u_L} \gamma_\mu U_L d_L
,\nn \\
&&\overline{u_R} \gamma_\mu d^0_R=\overline{u_R} \gamma_\mu U_R d_R.
\eea
The neutral currents keep their diagonal form as
\bea
\overline{u_L} \gamma_\mu u_L,\quad \overline{u_R} \gamma_\mu u_R,\quad 
\overline{d_L} \gamma_\mu d_L,\quad \overline{d_R} \gamma_\mu d_R.
\eea
In terms of the new basis, the downtype mass matrix ${\cal M_D}^0$
is changed into
\bea
{\cal M_D}&=&\begin{pmatrix}
U_L^\dagger & 0 \\
0 & 1
\end{pmatrix}
{\cal M_D}^0\begin{pmatrix}
U_R & 0 \\
0 & 1
\end{pmatrix}\nn\\
&=&\begin{pmatrix} 0 & y_{\Delta dL} v_L\\ 
                        y_{\Delta dR}^\dagger v_R & D_D \end{pmatrix}.
\label{eq:Bigd}
\eea
Note that in the new basis, the downtype Yukawa matrices 
are given by the triangular matrices. 
To summarize, at this stage, the mass terms of the quarks are
\bea
&& \begin{pmatrix} \overline{u_L} & \overline{\U_L} \end{pmatrix}
{\cal M_U} \begin{pmatrix} u_R \\ \U_R \end{pmatrix}, \quad
\begin{pmatrix} \overline{d_L} & \overline{\D_L} \end{pmatrix}
{\cal M_D} \begin{pmatrix} d_R \\ \D_R \end{pmatrix}.
\eea

\section{Diagonalization of the Mass matrices}
In the previous section, we performed the unitary transformation on $SU(2)$
doublet fields. In this section, we carry out the diagonalization 
of the $ 6 \times 6$ mass matrices ${\cal M_U}$ and ${\cal M_D}$.
Therefore, by the unitary transformation, 
doublet and singlet quarks are mixed in the mass eigenstates.
Now we diagonalize the mass matrices
given in Eqs. (\ref{eq:Bigu}) and (\ref{eq:Bigd}),
\bea
V_{uL}^\dagger {\cal M_U} V_{uR}=
\begin{pmatrix} d_u& 0\\ 
                        0 & \tD_U \end{pmatrix}.
\label{eq:vulvur}
\eea
Note that $d_u$ is a diagonal mass matrix for light uptype quarks 
and $\tD_U$ denotes that for heavy quarks.
The downtype mass matrix is diagonalized as
\bea
V_{dL}^\dagger {\cal M_D} V_{dR}=
\begin{pmatrix} d_d& 0\\
                        0 & \tD_D \end{pmatrix},
\label{eq:vdlvdr}
\eea
where $d_d$ is a diagonal mass matrix for light 
downtype quarks and $\tD_D$ is that for heavy quarks.
In terms of the mass eigenstates, charged currents and neutral currents
are written as,
\bea
&&\overline{u_{Li}} \gamma_\mu U_{Lij} d_{Lj}= \V_{L\alpha\beta}\overline{u^m_{L\alpha}}\gamma_\mu
d^m_{L\beta},
\label{eq:leftcc} \\
&&\overline{u_{Ri}} \gamma_\mu U_{Rij} d_{Rj}= \V_{R\alpha\beta}\overline{u^m_{R\alpha}}\gamma_\mu
d^m_{R\beta},
\label{eq:rightcc} \\
&&\overline{u_{Li}} \gamma_\mu u_{Li}=\Z_{uL \alpha \beta} \overline{u^m_{L \alpha}}\gamma_\mu u^m_{L \beta}, 
\label{eq:ulnc} \\
&&\overline{u_{Ri}} \gamma_\mu u_{Ri}=\Z_{uR \alpha \beta} \overline{u^m_{R \alpha}}
\gamma_\mu u^m_{R \beta}, 
\label{eq:urnc} \\
&&\overline{d_{Li}} \gamma_\mu d_{Li}=\Z_{dL \alpha \beta} \overline{d^m_{L \alpha}}
\gamma_\mu d^m_{L \beta}, 
\label{eq:dlnc} \\
&&\overline{d_{Ri}} \gamma_\mu d_{Ri}=\Z_{dR \alpha \beta} \overline{d^m_{R \alpha}}
\gamma_\mu d^m_{R \beta}
\label{eq:drnc},
\eea
where $u^m_\alpha$ and $d^m_\alpha (\alpha=1,\cdots,6)$ denote the mass eigenstates.

We parametrize $V_{qL}$ and $V_{qR}$ $(q=u,d)$ with $3 \times 3$ submatrices
as
\bea
V_{qL} =\begin{pmatrix} K_{qL} & R_{qL} \\
                     S_{qL} & T_{QL} \end{pmatrix},\quad
V_{qR} =\begin{pmatrix} K_{qR} & R_{qR} \\
                     S_{qR} & T_{QR} \end{pmatrix}.
\label{eq:vqlvqr}
\eea
The $6\times 6$ mixing matrices $\V_L$ and $\V_R$ for the charged currents
in Eqs. (\ref{eq:leftcc}) and
(\ref{eq:rightcc}) are written as
\bea
\V_L&=&
\begin{pmatrix}
K_{uL}^\dagger U_LK_{dL} & K_{uL}^\dagger U_LR_{dL}\\
R_{uL}^\dagger U_LK_{dL} & R_{uL}^\dagger U_LR_{dL}
\end{pmatrix}, \\
\V_R&=&
\begin{pmatrix}
K_{uR}^\dagger U_RK_{dR} & K_{uR}^\dagger U_RR_{dR}\\
R_{uR}^\dagger U_RK_{dR} & R_{uR}^\dagger U_RR_{dR}
\end{pmatrix}.
\eea
The mixing matrices for the neutral currents in Eqs.(\ref{eq:ulnc})-(\ref{eq:drnc}) are given by
\bea
\Z_{uL}&=&\begin{pmatrix}
K_{uL}^\dagger K_{uL} &  K_{uL}^\dagger R_{uL}\\
R_{uL}^\dagger K_{uL} & R_{uL}^\dagger R_{uL}
\end{pmatrix},\quad
\Z_{uR}=\begin{pmatrix}
K_{uR}^\dagger K_{uR} &  K_{uR}^\dagger R_{uR}\\
R_{uR}^\dagger K_{uR} & R_{uR}^\dagger R_{uR}
\end{pmatrix},\label{eq:Zu}\\
\Z_{dL}&=&\begin{pmatrix}
K_{dL}^\dagger K_{dL} &  K_{dL}^\dagger R_{dL}\\
R_{dL}^\dagger K_{dL} & R_{dL}^\dagger R_{dL}
\end{pmatrix},\quad
\Z_{dR}=\begin{pmatrix}
K_{dR}^\dagger K_{dR} &  K_{dR}^\dagger R_{dR}\\
R_{dR}^\dagger K_{dR} & R_{dR}^\dagger R_{dR}
\end{pmatrix}.\label{eq:Zd}
\eea
The quark interaction terms induced by neutral currents are written in terms of mass eigenstate quark fields and mass eigenstate 
gauge fields as follows:
\bea
-\mathcal{L}_{\rm NC}&=&+\frac{2}{3}e\bar{u}_\alpha^m\sla{A}u^m_\alpha
-\frac{1}{3}e\bar{d}_\alpha^m\sla{A}d_\alpha^m \nn \\
&-&\frac{2}{3}g_1(c_Rs_Wc_\xi+s_Rs_\xi)
\bar{u}_{\alpha}^m\sla{Z}u_{\alpha}^m
+\frac{1}{3}g_1(c_Rs_Wc_\xi+s_Rs_\xi)
\bar{d}_{\alpha}^m\sla{Z}d_{\alpha}^m
\nn \\
&+&\frac{2}{3}g_1(s_Wc_Rs_\xi-s_Rc_\xi)\bar{u}_{\alpha}^m
\sla{Z}^\prime u_{\alpha}^m 
-\frac{1}{3}g_1(s_Wc_Rs_\xi-s_Rc_\xi)\bar{d}_{\alpha}^m
\sla{Z}^\prime d_{\alpha}^m  \nn \\
&+&\left[\frac{1}{2}\Z_{uL\alpha\beta}\left(g_Lc_Wc_\xi+g_1\left(c_Rs_Wc_\xi+s_Rs_\xi\right)\right)\right]\bar{u}_{L\alpha}^m\sla{Z}u_{L\beta}^m\nn\\
&+&\left[-\frac{1}{2}\Z_{dL\alpha\beta}\left(g_Lc_Wc_\xi+g_1(c_Rs_Wc_\xi+s_Rs_\xi)\right)\right]\bar{d}_{L\alpha}^m\sla{Z}d_{L\beta}^m\nn\\
&+&\left[\frac{1}{2}\Z_{uR\alpha\beta}\left(g_R(c_Rs_\xi-s_Rs_Wc_\xi)+g_1(c_Rs_Wc_\xi+s_Rs_\xi)\right)\right]\bar{u}_{R\alpha}^m\sla{Z}u_{R\beta}^m\nn\\
&+&\left[-\frac{1}{2}\Z_{dR\alpha\beta}\left(g_R(c_Rs_\xi-s_Rs_Wc_\xi)+g_1(c_Rs_Wc_\xi+s_Rs_\xi)\right)\right]\bar{d}_{R\alpha}^m\sla{Z}d_{R\beta}^m\nn\\
&+&\left[\frac{1}{2}\Z_{uL\alpha\beta}\left(-g_Lc_Ws_\xi+g_1(s_Rc_\xi-c_Rs_Ws_\xi)\right)\right]\bar{u}_{L\alpha}^m\sla{Z}^\prime u_{L\beta}^m\nn\\
&+&\left[-\frac{1}{2}\Z_{dL\alpha\beta}\left(-g_Lc_Ws_\xi+g_1(s_Rc_\xi-c_Rs_Ws_\xi)\right)\right]\bar{d}_{L\alpha}^m\sla{Z}^\prime d_{L\beta}^m\nn\\
&+&\left[\frac{1}{2}\Z_{uR\alpha\beta}\left(g_R(c_Rc_\xi+s_Rs_Ws_\xi)+g_1(s_Rc_\xi-c_Rs_Ws_\xi)\right)\right]\bar{u}_{R\alpha}^m\sla{Z}^\prime u_{R\beta}^m\nn\\
&+&\left[-\frac{1}{2}\Z_{dR\alpha\beta}\left(g_R(c_Rc_\xi+s_Rs_Ws_\xi)+g_1(s_Rc_\xi-c_Rs_Ws_\xi)\right)\right]\bar{d}_{R\alpha}^m\sla{Z}^\prime d_{R\beta}^m.\nn\\
\label{NCofquark}
\eea
In Eq.(\ref{NCofquark}), we used the 
notation of the mass eigenstates of neutral gauge fields,
\bea
\begin{pmatrix}
A\\
Z\\
Z^\prime
\end{pmatrix}
=
\begin{pmatrix}
c_Wc_R & s_W & c_Ws_R\\
-(s_Wc_Rc_\xi+s_Rs_\xi) & c_Wc_\xi & c_Rs_\xi-s_Ws_Rc_\xi\\
s_Wc_Rs_\xi-s_Rc_\xi & -c_Ws_\xi & c_Rc_\xi+s_Ws_Rs_\xi
\end{pmatrix}
\begin{pmatrix}
B\\
W_L^3\\
W_R^3
\end{pmatrix},
\label{gaugemat}
\eea
where the mixing angles of the neutral gauge bosons satisfy the following
equations.
\bea
&g_Lt_W=g_1c_R,&\nn\\
&g_Rt_R=g_1,&\nn\\
&\tan 2 \xi=-\displaystyle\frac
{s_R^2\sin 2\theta_Rv_L^2}
{s_W\left[v_R^2
+\left(s_R^4-\frac{\sin^2 2\theta_R}{\sin^2 2\theta_W}\right)v_L^2\right]}
\eea
They are derived by taking the vacuum expectation value of
the bidoublet Higgs field zero in the formulas of Ref. \cite{Chay:1998hd}.
In order to acquire the derivation of the relation(\ref{gaugemat}) and the definition of mixing parameters $s_W,s_R,s_\xi,$ and $e$, see Ref. 
\cite{Chay:1998hd}.
\section{The approximate formulas for the mixing matrices}
So far, we derive the exact formulas for the mixing matrices.
In this section, we carry out the diagonalization
of the mass matrices and determine the unitary matrices for the
diagonalization.  In Appendix B, we show the procedure of the
diagonalization and the approximation.
We have determined the submatrices of the unitary matrices
$V_{qL}$ and $V_{qR}$ in
Eqs.(\ref{eq:vulvur}), (\ref{eq:vdlvdr}), and (\ref{eq:vqlvqr}).
The approximate formulas on
$K_{uL}$ in Eq.(\ref{eq:akul}) and $R_{uL}$ in Eq.(\ref{eq:arul})
are given as
\bea
K_{uL}&=&\begin{pmatrix} 1 & 
\frac{M_C y_{u L1}y_{u R 21}^\ast}{M_U y_{u L 2}y_{u R 2}} & 
\frac{M_T y_{u L1}y_{u R 31}^\ast}{M_U y_{u L 3}y_{u R 3}} \\
-\frac{M_C y_{u L1}y_{u R 21}}{M_U y_{u L 2}y_{u R 2}} & 1 & 
\frac{M_T y_{u L2}y_{u R 32}^\ast}{M_C y_{u L 3}y_{u R 3}} \\
\frac{M_T}{M_U}
\frac{y_{uL1}(y_{uR 32} y_{uR21}-y_{uR2}y_{uR31})}
{y_{uR2}y_{uL3} y_{uR3}}
&  -\frac{M_T y_{u L2}y_{u R 32}}{M_C y_{u L3}y_{u R 3}} & 1 
\end{pmatrix},
\label{eq:kul} \\
R_{uL}&=&
\begin{pmatrix}
\frac{v_L}{M_U}y_{uL1} & 0 & 0  \\
\frac{v_L}{M_U}y_{uL21} &\frac{v_L}{M_C}y_{uL2}&0\\
\frac{v_L}{M_U}y_{uL31}&\frac{v_L}{M_C}y_{uL32}
&\frac{v_LM_T}{D_{T}^2}y_{uL3}
\end{pmatrix},
\label{eq:rul}
\eea
where $D_T$ denotes the mass eigenvalue of the
lightest state of the 
heavy uptype quarks and the definition can be found in
Eq.(\ref{eq:d033}). 
Similarly, the downtype mixing matrices $K_{dL}$, $R_{dL}$ have following forms:
\bea
K_{dL}&=&\begin{pmatrix} 1 & 
\frac{M_S y_{d L1}y_{d R 21}^\ast}{M_D y_{d L 2}y_{d R 2}} & 
\frac{M_B y_{d L1}y_{d R 31}^\ast}{M_D y_{d L 3}y_{d R 3}} \\
-\frac{M_S y_{d L1}y_{d R 21}}{M_D y_{d L 2}y_{d R 2}} & 1 & 
\frac{M_B y_{d L2}y_{d R 32}^\ast}{M_S y_{d L 3}y_{d R 3}} \\
\frac{M_B}{M_D}
\frac{y_{dL1}(y_{dR 32}y_{dR21}-y_{dR2}y_{dR31})}
{y_{dR2}y_{dL3} y_{dR3}}&
-\frac{M_B y_{d L2}y_{d R 32}}{M_S y_{d L3}y_{d R 3}} & 1 
\end{pmatrix},
\label{eq:akdl}\\
R_{dL}&=&
\begin{pmatrix}
\frac{v_L}{M_D}y_{dL1} & 0 & 0  \\
\frac{v_L}{M_D}y_{dL21} &\frac{v_L}{M_S}y_{dL2}&0\\
\frac{v_L}{M_D}y_{dL31}&\frac{v_L}{M_S}y_{dL32}&\frac{v_L}{M_B}y_{dL3}
\end{pmatrix}.
\label{eq:ardl}
\eea 
The approximate forms for 
$K_{uR}$, $R_{uR}$, $K_{dR}$, and $R_{dR}$ are also derived 
using the formulas
\bea
K_{qR}=y_{qR}v_RS_{qL}/d_q,\quad
R_{qR}=y_{qR}v_RT_{QL}/\tD_Q,
\label{eq:relations}
\eea
where Eq.(\ref{eq:relations}) is derived using Eq.(\ref{eq;16}).
By substituting the approximate formulas for 
$S_{qL}$ and $T_{QL}$ given in Eqs.(\ref{eq:asul}) and (\ref{eq:TuL}),
$K_{qR}$ and $R_{qR}$ are given as
\bea
K_{uR}&=&-y_{\Delta uR}\frac{D_U}{D_{0U}^2}
y_{\Delta uL}^\dagger
K_{uL}\frac{v_Lv_R}{d_u}\nn\\
&=&	-\begin{pmatrix}
			1 & \frac{M_C}{M_U}\frac{y_{uR1}y_{uL21}^*}								{y_{uL2}y_{uR2}}
			& \frac{D_{T}}{M_U}\frac{y_{uR1}y_{uL31}^*}{y_{uL3}						y_{uR3}}\\
			\frac{M_T}{M_C}\frac{y_{uL32}^*(y_{uR21}y_{uR32}-						y_{uR2}y_{uR31})}{y_{uR1}y_{uL3}y_{uR3}}
			-\frac{M_C}{M_U}\frac{y_{uR21}y_{uR21}y_{uL21}^*}
			{y_{uR1}y_{uL2}y_{uR2}}
			& 1
			& \frac{D_{T}}{M_C}\frac{y_{uR2}y_{uL32}^*}							{y_{uL3}y_{uR3}}\\
			(1-\frac{M_T^2}{D_{T}^2})\frac{y_{uR2}y_{uR31}
			-y_{uR21}y_{uR32}}{y_{uR1}y_{uR2}}
			& (1-\frac{M_T^2}{D_{T}^2})\frac{y_{uR32}}{y_{uR2}}
			& \frac{M_T}{D_{T}}+\frac{D_{T}}{M_C}
			\frac{y_{uL32}^*y_{uR32}}{y_{uL3}y_{uR3}}
\end{pmatrix},\nn\\
\label{eq:kur} \\
R_{uR}&=&y_{\Delta uR}\frac{v_R}{D_{0U}}\nn \\
&=&
\begin{pmatrix}
\frac{v_R}{M_U}y_{uR1} & 0 & 0\\
\frac{v_R}{M_U}y_{uR21} &\frac{v_R}{M_C}y_{uR2}  &0\\
\frac{v_R}{M_U}y_{uR31} & \frac{v_R}{M_C}y_{uR32} & \frac{v_R}{D_{T}}y_{uR3}
\end{pmatrix},
\label{eq:Rur}\\
K_{dR}&=&-y_{\Delta dR}\frac{1}{D_D}y_{\Delta dL}^\dagger
K_{dL}\frac{v_Lv_R}{d_d}\nn\\
&=&	-\begin{pmatrix}
			1 &
			\frac{M_S}{M_D}\frac{y_{dR1}y_{dL21}^*}
			{y_{L2}y_{dR2}} &
			\frac{M_B}{M_D}\frac{y_{dR1}y_{dL31}^*}									{y_{dL3}y_{dR3}} \\
			-K_{dR21} &
			1 &
			\frac{M_B}{M_S}\frac{y_{dR2}y_{dL32}^*}									{y_{dL3}y_{dR3}} \\
			-K_{dR31} &
			\frac{M_S}{M_D}\frac{y_{dL21}^*y_{dR31}}
			{y_{dL2}y_{dR2}}-\frac{M_B}{M_S}\frac{y_{dL32}^*
			y_{dR32}y_{dR32}}{y_{dR2}y_{dL3}y_{dR3}} &
			1
\end{pmatrix},
\label{eq:Kdr} \\
&&K_{dR21}=-\frac{M_B}{M_S}
			\frac{y_{dL32}^*(y_{dR21}y_{dR32}-y_{dR2}y_{dR31})}
			{y_{dR1}y_{dL3}y_{dR3}}
			+\frac{M_S}{M_D}\frac{y_{dL21}^*y_{dR21}y_{dR21}}
			{y_{dR1}y_{dL2}y_{dR2}}
,\nn\\
&&K_{dR31}=-\frac{M_B}{M_S}
			\frac{y_{dL32}^*(y_{dR21}y_{dR32}y_{dR32}
			-y_{dR2}y_{dR31}y_{dR32})}
			{y_{dR1}y_{dR2}y_{dL3}y_{dR3}}
			+\frac{M_S}{M_D}\frac{y_{dL21}^*y_{dR21}y_{dR31}}
			{y_{dR1}y_{dL2}y_{dR2}}
,\nn\\
R_{dR}&=&y_{\Delta dR}\frac{v_R}{D_{D}}\nn \\
&=&
\begin{pmatrix}
\frac{v_R}{M_D}y_{dR1} & 0 & 0\\
\frac{v_R}{M_D}y_{dR21} &\frac{v_R}{M_S}y_{dR2}  &0\\
\frac{v_R}{M_D}y_{dR31} & \frac{v_R}{M_S}y_{dR32} & \frac{v_R}{M_B}y_{dR3}
\end{pmatrix},
\label{eq:ardr}
\eea
where the definition of $D_{0U}$ can be found in Eq.(\ref{eq:d0u}).

We summarize the results of the mixing matrices.
The left-handed charged current, $\V_L$ is determined in a good approximation as follows:
\bea
\V_L \simeq \begin{pmatrix} U_L & U_L R_{uL} \\ R_{dL}^\dagger U_L & R_{dL}^\dagger U_L R_{uL}
\end{pmatrix}.
\eea
where we ignore the corrections suppressed by heavy quarks masses
by setting $K_{uL} \simeq K_{dL} \simeq K_{dR} \simeq 1$.
The $3 \times 3$ submatrix 
which corresponds to light quark mixings is mostly determined by the $3 \times 3$
unitary matrix $U_L$. In our parametrization $U_L$ includes
five CP violating phases $\alpha_{L1}, \alpha_{L1}, \delta_L, \beta_{L1},\beta_{L2}$.
The mixing between the light quark and heavy quark is suppressed
by a factor of $\frac{v_L}{D_{0Uii}}\ll1$ or $\frac{v_L}{D_{Di}}\ll1$.
The mixing among heavy quarks is suppressed by a factor
of the product $\frac{v_L^2}{D_{0Uii}D_{Dj}}$. 
One finds that the mixing of  the heavy uptype quarks and the  light downtype
quarks corresponding to ${\cal V}_{R6i}$
 ($i = 1- 3$) is large. The large mixing occurs
because the component of
$R_{u R 33}$ is not suppressed. This phenomenon is related to the enhancement 
mechanism of the top quark mass as shown in 
Refs. [8-9]

The flavor changing neutral current (FCNC) for up quarks is determined by $Z_{uL},Z_{uR}$ in Eq.(\ref{eq:Zu}).
We first show the approximate formulas for 
the FCNC among the light uptype quarks,
$Z_{uL ij}$ $(i,j=1-3)$. They are derived using the relation,
 $\Z_{uLij}=(K_{uL}^\dagger K_{uL})_{ij}
\simeq \delta_{ij}-(S^\dagger_{uL}S_{uL})_{ij}$,
\bea
&&\Z_{uL11}\simeq 1-\left(\displaystyle\frac{m_u}{v_R}\right)^2
\left[\displaystyle\sum_{k=1}^2(y_{\Delta uR}^{-1})_{k1}^*
(y_{\Delta uR}^{-1})_{k1}+\left(\frac{M_T}{D_{T}}\right)^4
|(y_{\Delta uR}^{-1})_{31}|^2\right],\nn\\
&&\Z_{uL12}\simeq -\displaystyle\frac{m_um_c}{v_R^2}\left[(y_{\Delta
uR}^{-1})_{22}(y_{\Delta uR}^{-1})_{21}^*
+\left(\frac{M_T}{D_{T}}\right)^4(y_{\Delta uR}^{-1})_{31}
(y_{\Delta uR}^{-1})_{32}
\right],\nn\\
&&\Z_{uL13}\simeq \displaystyle\frac{m_um_c}{v_R^2}
\left(\frac{1}{y_{uL2}}(y_{\Delta uR}^{-1})^*_{21}
(y_{\Delta u R}^{-1})_{32}(y_{\Delta u R}^{-1})^*_{33}
\right)-\frac{m_um_t}{v_R^2}
\left(\frac{M_T}{D_{T}}\right)^3
(y_{\Delta uR}^{-1})_{31}(y_{\Delta uR}^{-1})_{33},\nn\\
&&\Z_{uL22}\simeq 1-
\left(\displaystyle\frac{m_c}{v_R}\right)^2
\left[(y_{\Delta uR}^{-1})_{22}^2+
\left(\displaystyle\frac{M_T}{D_{T}}\right)^4|
(y_{\Delta u R}^{-1})_{31}|^2\right],\nn\\
&&\Z_{uL23}\simeq -\left(\displaystyle\frac{m_c}{v_R}\right)^2
\displaystyle\frac{y_{uL32}^*}{y_{uR2}^2y_{uL2}}
-\frac{m_cm_t}{v_R^2}\left(\frac{M_T}{D_{T}}\right)^3
(y_{\Delta u R}^{-1})^*_{32}(y_{\Delta u R}^{-1})_{33}^{-1}
,\nn\\
&&\Z_{uL33}\simeq 1-\left(\displaystyle\frac{m_c}{v_R}\right)^2
\displaystyle\frac{|y_{uL32}|^2}{y_{uL2}^2y_{uR2}^2}
-\left(\displaystyle\frac{m_t}{v_R}\right)^2
\left(\displaystyle\frac{M_T}{D_{T}}\right)
(y_{\Delta u R}^{-1})_{33}^2,
\eea
We note that the CP violation of the tree-level FCNC for the left-handed
current is determined by the CP violating phases in the right-handed 
Yukawa couplings $y_{u R ij} (i>j)$ and left-handed Yukawa
coupling $y_{uL 32}$. The strength of 
the FCNC is naturally suppressed by the $SU(2)_R$ breaking scale.
The FCNC of the left-handed current 
between light uptype quarks and the heavy ones can be written as 
\bea
K_{uL}^\dagger R_{uL}\simeq
\begin{pmatrix}
\frac{m_u}{v_R}\frac{1}{y_{uR1}}&
-\frac{m_u}{v_R}\frac{y_{uR21}^*}{y_{uR1}y_{uR2}}&
(\frac{M_T}{D_{T}})^2\frac{m_u}{v_R}
\frac{y_{uR21}^*y_{uR32}^*-y_{uR2}y_{uR31}^*}{y_{uR1}y_{uR2}y_{uR3}}\\
\frac{m_u}{v_R}\frac{y_{uL21}}{y_{uR1}y_{uL1}} &
\frac{m_c}{v_R}\frac{1}{y_{uR2}}&
-(\frac{M_T}{D_{T}})^2\frac{m_c}{v_R}\frac{y_{uR32}^*}{y_{uR2}y_{uR3}}\\
\frac{m_u}{v_R}\frac{y_{uL31}}{y_{uL1}y_{uR1}} &
\frac{m_c}{v_R}\frac{y_{uL32}}{y_{uL2}y_{uR2}} &
(\frac{M_T}{D_{T}})\frac{m_t}{v_R}\frac{1}{y_{uR3}}
\end{pmatrix}.
\eea
The strength of the FCNC between the heavy left-handed uptype quark and 
light uptype quark is suppressed by the $SU(2)_R$ breaking scale.
We also note that CP violation is determined by the phases of $y_{L32}$ and
$y_{R ij}, (i>j)$. 
The FCNC of the left-handed current 
among the heavy uptype quarks is given as
\bea
R_{uL}^\dagger R_{uL}\simeq
\begin{pmatrix}
(\frac{m_u}{v_R})^2\frac{y_{uL1}^2+|y_{uL21}|^2+|y_{uL31}|^2}{y_{uL1}^2y_{uR1}^2}&
\frac{m_um_c}{v_R^2}\frac{y_{uL2}y_{uL21}^*+y_{uL31}^*y_{uL32}}{y_{uL1}y_{uL2}y_{uR1}y_{uR2}}&
\frac{M_T}{D_{T}}\frac{m_um_t}{v_R^2}\frac{y_{uL31}^*}{y_{uL1}y_{uR1}y_{uR3}}\\
\frac{m_um_c}{v_R^2}\frac{y_{uL2}y_{uL21}+y_{uL31}y_{uL32}^\ast}
{y_{uL1}y_{uL2}y_{uR1}y_{uR2}}&
(\frac{m_c}{v_R})^2\frac{y_{uL2}^2+|y_{uL32}|^2}{y_{uL2}^2y_{uR2}^2}&
\frac{M_T}{D_{T}}\frac{m_cm_t}{v_R^2}\frac{y_{uL32}^*}{y_{uL2}y_{uR2}y_{uR3}}\\
\frac{M_T}{D_{T}}\frac{m_um_t}{v_R^2}\frac{y_{uL31}}{y_{uL1}y_{uR1}y_{uR3}}&
\frac{M_T}{D_{T}}\frac{m_cm_t}{v_R^2}\frac{y_{uL32}}{y_{uL2}y_{uR2}y_{uR3}}
&
(\frac{M_T}{D_{T}})^2\frac{m_t^2}{v_R^2}\frac{1}{y_{uR3}^2}
\end{pmatrix}.\nn \\
\eea
All the components are suppressed by a factor of $\frac{1}{v_R^2}$.
The CP violation of the FCNC is determined by the left-handed Yukawa coupling
 $y_{uL 32}$.
Similarly, the FCNCs for the light right-handed uptype quarks are given as
\bea
\Z_{uRij}&=&(K_{uR}^\dagger K_{uR})_{ij}\simeq\delta_{ij}-(S_{uR}^\dagger
S_{uR})_{ij} \nn \\
&\simeq& \delta_{ij}-\frac{m_u^im_u^j}{v_L^2}\displaystyle\sum_{k\geq i,j}
(y^{-1}_{\Delta uL})^*_{ki}(y^{-1}_{\Delta 
uL})_{kj},\quad ({\rm for}\:i,j=1,2,3).
\label{eq:ZuRij}
\eea
Note that the flavor diagonal coupling $Z_{uR33}$ of the right-handed top quark
current
$\overline{t^m_R} \gamma_\mu t^m_R$ is suppressed,
\bea
Z_{uR 33} \simeq 1-\frac{m_t^2}{y_{uL3}^2 v_L^2}.
\label{eq:ZuR33}
\eea 
The suppression of the FCNC for the right-handed current is weaker than
that of the left-handed one.  It is suppressed by a factor of $\frac{1}{v_L^2}$. The CP violation of the FCNC in Eq.(\ref{eq:ZuRij})
is determined by a phase of $y_{uL32}$. We note that
the same phase appears in the FCNC of the left-handed current among the heavy uptype quarks.  
Below we show all the components of the FCNC couplings for the right-handed currents between the light uptype quark and the heavy uptype quark:
\bea
&&\Z_{uR14}=-\frac{m_u}{v_L}\frac{1}{y_{uL1}y_{uR1}^2y_{uR2}}
\left[y_{uR1}^2y_{uR2}
\right.\nn\\
&&\qquad\qquad+\left.\left(1-\left(\frac{m_tM_T}{v_Lv_R}\right)^2
\frac{1}{y_{uL3}^2y_{uR3}^2}\right)(y_{uR2}|y_{uR31}|^2
-y_{uR21}^*y_{uR31}y_{uR32}^*)\right]
,\nn\\
&&\Z_{uR15}=\frac{m_c}{v_L}\left[1-\left(\frac{m_tM_T}{v_Lv_R}\right)^2\frac{1}{y_{uL3}^2y_{uR3}^2}\right]
\frac{y_{uR32}(y_{uR21}^*y_{uR32}^*-y_{uR2}y_{uR31}^*)}
{y_{uL2}y_{uR1}y_{uR2}^2},\nn\\
&&\Z_{uR16}=\frac{m_t}{v_L}\left[1-\left(\frac{m_tM_T}{v_Lv_R}\right)^2\frac{1}{y_{uL3}^2y_{uR3}^2}\right]
\frac{y_{uR21}^*y_{uR32}^*-y_{uR2}y_{uR31}^*}
{y_{uR1}y_{uR2}y_{uL3}},\nn\\
&&\Z_{uR24}=-\frac{m_u}{v_L}\frac{1}{y_{uL1}y_{uR1}y_{uR2}}\left[y_{uR2}y_{uR21}
+\left(1-\left(\frac{m_tM_T}{v_Lv_R}\right)^2
\frac{1}{y_{uL3}^2y_{uR3}^2}\right)
y_{uR31}y_{uR32}^*\right],\nn\\
&&\Z_{uR25}=-\frac{m_c}{v_L}\frac{1}{y_{uL2}y_{uR2}^2}
\left[y_{uR2}^2+\left(1-\left(\frac{m_tM_T}{v_Lv_R}
\right)^2\frac{1}{y_{uL3}^2y_{uR3}^2}
\right)|y_{uR32}|^2\right],\nn\\
&&\Z_{uR26}=-\frac{m_t}{v_L}\frac{y_{uR32}^*}{y_{uR2}y_{uL3}}
\left[1-\left(\frac{m_tM_T}{v_Lv_R}\right)^2\frac{1}
{y_{uL3}^2y_{uR3}^2}
\right],\nn\\
&&\Z_{uR34}=-\frac{m_u}{v_L}\frac{1}{y_{uL1}y_{uR1}}
\left[\frac{m_c}{m_t}\frac{y_{uL32}(y_{uR2}y_{uR21}+
y_{R31}y_{R32}^*)}{y_{uL2}y_{uR2}}
+\frac{m_tM_T}{v_Lv_R}\frac{y_{uR31}}{y_{uL3}y_{uR3}}
\right],\nn\\
&&\Z_{uR35}=-\frac{m_c}{v_L}\frac{1}{y_{uL2}y_{uR2}}
\left[\frac{m_c}{m_t}\frac{y_{uL32}(y_{uR2}^2+|y_{uR32}|^2)}
{y_{uL2}y_{uR2}}+\frac{m_tM_T}{v_Lv_R}\frac{y_{uR32}}
{y_{uL3}y_{uR3}}
\right],\nn\\
&&\Z_{uR36}=-\frac{m_t}{v_L}\frac{1}{y_{uL3}}
\left[\frac{m_c}{m_t}\frac{y_{uL32}y_{uR32}^*}
{y_{uL2}y_{uR2}}+\frac{m_tM_T}{v_Lv_R}\frac{1}
{y_{uL3}y_{uR3}}
\right].
\label{eq:ZRlh}
\eea
We observe that CP violation is determined by the four phases,
${\rm Im}(y_{uRij})$ $(i>j)$ and ${\rm Im}(y_{uL32})$.
The FCNC among the heavy right-handed uptype quarks is given as
\bea
\Z_{uR i+3,j+3}\simeq 
(R_{uR}^\dagger R_{uR})_{ij}\simeq\frac{m_u^im_u^j}{v_L^2}
\frac{1}{y_{uLi}y_{uRi}y_{uLj}y_{uRj}}\displaystyle\sum_{k\geq i,j}^3y_{uRki}^*y_{uRkj}.
\eea
Note that the flavor diagonal coupling $\Z_{uR 66}
\simeq \frac{m_t^2}{y_{uL3}^2 v_L^2} $ is 
not suppressed, which is in contrast to the coupling $Z_{uR 33}$ in 
Eq.(\ref{eq:ZuR33}).

Next we show the FCNC of the down-quark sector in 
Eq.(\ref{eq:Zd}). The approximate formulas for the FCNC for the left-handed currents is given by
\bea
&&\Z_{dLij}=\delta_{ij}-\frac{m_d^im_d^j}{v_R^2}\displaystyle\sum_{k\geq i,j}^3
(y^{-1}_{\Delta dR})^*_{ki}(y^{-1}_{\Delta dR})_{kj}
,\quad ({\rm for}\:i,j=1,2,3),\nn \\
&& K_{dL}^\dagger R_{dL}\simeq\begin{pmatrix}
\frac{m_d}{v_R}\frac{1}{y_{dR1}}&
-\frac{m_d}{v_R}\frac{y_{dR21}^*}{y_{dR1}y_{dR2}}&
\frac{m_d}{v_R}\frac{y_{dR21}^*y_{dR32}^*-y_{dR2}y_{dR31}^*}{y_{dR1}y_{dR2}y_{dR3}}\\
\frac{m_d}{v_R}\frac{y_{dL21}}{y_{dL1}y_{dR1}}&
\frac{m_s}{v_R}\frac{1}{y_{dR2}}&
-\frac{m_s}{v_R}\frac{y_{dR32}^*}{y_{dR2}y_{dR3}}\\
\frac{m_d}{v_R}\frac{y_{dL31}}{y_{dL1}y_{dR1}}&
\frac{m_s}{v_R}\frac{y_{dL32}}{y_{dL2}y_{dR2}}&
\frac{m_b}{v_R}\frac{1}{y_{dR3}}
\end{pmatrix},\\
&& (R_{dL}^\dagger R_{dL})_{ij}\simeq
\frac{m_d^im_d^j}{v_R^2}\frac{1}{y_{dLi}y_{dR i}
y_{dL j}y_{dRj}}\displaystyle\sum_{k\geq i,j}^3
y_{dLki}^*y_{dLkj}, \quad ({\rm for}\:i,j=1,2,3).\nn
\label{eq:fcncdL}
\eea
As we can easily see from Eq.(\ref{eq:fcncdL}), the FCNC of the down-quark sector is much simpler than that of the up-quark sector.
The FCNC for the left-handed current among the light down-type quarks
is suppressed by a factor of $\frac{1}{v_R^2}$. The same suppression occurs in
the FCNC among heavy quarks. The FCNC between 
the heavy quark
and light quark is suppressed by a factor of $\frac{1}{v_R}$.
For the right-handed current, the FCNC couplings  are given as
\bea
&&\Z_{dRij}=(K_{dR}^\dagger K_{dR})_{ij}\simeq\delta_{ij}-(S_{dR}^\dagger
S_{dR})_{ij}\simeq \delta_{ij}-
\frac{m_d^im_d^j}{v_L^2}\displaystyle\sum_{k\geq i,j}^3
(y^{-1}_{\Delta dL})^*_{ki}(y^{-1}_{\Delta dL})_{kj}, ({\rm for}\:i,j=1,2,3), \nn \\
&&\Z_{dR14}=-\frac{m_d}{v_L}\frac{1}{y_{dL1}},\nn\\
&&\Z_{dR15}=\frac{m_d}{v_L}\frac{y_{dL21}y_{dR21}^*(y_{dR2}y_{dR21}^*
+y_{dR31}^*y_{dR32})}
{y_{dL1}y_{dR1}^2y_{dL2}y_{dR2}}\nn\\
&&\qquad\qquad+\frac{m_s^2}{v_Lm_b}
\frac{y_{dL32}[y_{dR2}^2(y_{dR2}y_{dR31}^*-y_{dR21}^*y_{dR32}^*)
+|y_{dR32}|^2(y_{dR2}y_{dR31}^*-y_{dR21}^*y_{dR32}^*)]}
{y_{dR1}y_{dL2}^2y_{dR2}^3},\nn\\
&&\Z_{dR16}=\frac{m_s}{v_L}\frac{y_{dL32}y_{dR32}^*
(y_{dR2}y_{dR31}^*-y_{dR21}^*y_{dR32}^*)}
{y_{dR1}y_{dL2}y_{dR2}^2y_{dL3}}+
\frac{m_dm_b}{m_sv_L}\frac{y_{dL21}y_{dR21}^*
y_{dR31}^*}{y_{dL1}y_{dR1}^2y_{dL3}},\nn^\\
&&\Z_{dR24}=-\frac{m_d}{v_L}\frac{y_{dR21}}{y_{dL1}y_{dR1}},\nn\\
&&\Z_{dR25}=-\frac{m_s}{v_L}\frac{1}{y_{dL2}}
,\nn\\
&&\Z_{dR26}=\frac{m_s}
{v_L}\frac{y_{dL32}y_{dR32}^*y_{dR32}^*}
{y_{dL2}y_{dR2}^2y_{dL3}}
-\frac{m_dm_b}{m_sv_L}\frac{y_{dL21}y_{dR31}^*}
{y_{dL1}y_{dR1}y_{dL3}},\nn\\
&&\Z_{dR34}=-\frac{m_d}{v_L}\frac{y_{dR31}}{y_{dL1}y_{dR1}},\nn\\
&&\Z_{dR35}=-\frac{m_s}{v_L}\frac{y_{dR32}}{y_{dL2}y_{dR2}},\nn\\
&&\Z_{dR36}=-\frac{m_b}{v_L}\frac{1}{y_{dL3}}, \\
&& (R_{dR}^\dagger R_{dR})_{ij}\simeq
\frac{m_d^im_d^j}{v_L^2}\frac{1}{y_{dLi}y_{dRi}y_{dLj}y_{dRj}}\displaystyle\sum_{k\geq i,j}^3
y_{dRki}^*y_{dRkj}.\nn
\eea
The FCNC among the light downtype quarks is suppressed by a factor of
$\frac{m_{d i} m_{d j}}{v_L^2}$.  Since
the downtype quarks' masses are smaller than 
$v_L$, the FCNC
for down-quark sector is naturally suppressed. We observe that 
the suppression of the FCNC among the heavy quarks 
is similar to the light-quark case. The FCNC from heavy quarks to
light quarks is also suppressed
by a factor of $\frac{1}{v_L}$, which is weaker than that for the left-handed
current. However, it is much suppressed compared with that of the corresponding
up-quark case.
Concerning CP violation, we observe that the CP violation of
the tree-level FCNCs are determined by the imaginary parts of the
triangular matrices of the Yukawa couplings $y_{\Delta}$. 
\section{Conclusion}
We study CP violation and flavor mixings in the quark sector of 
the universal seesaw model.
We find the number of independent parameters in a specific weak basis.
The basis is obtained using all the freedom of the WBT.
There is no redundancy due to WBT in the parameters left. Therefore, the
number of the parameters in such weak basis 
corresponds to the number of independent
parameters of the model.
The results of the number of parameters (real parts and imaginary parts)
are summarized in Table I and in Table II for the case where the singlet quark generation number 
is identical to the doublet quark generation number. 
The number of CP violating parameters is also obtained
by counting the number of CP invariant conditions that are nontrivially satisfied,
which agrees with the one in the specific weak basis. 

For the three-generation model, the number 
of CP violating phases is 19.
The corresponding  CP violating WB invariants are constructed in terms
of the Yukawa matrices and singlet quark matrices. 
To identify the CP violation and mixings in mass eigenstates of quarks,
we study the unitary matrices $V_{qL}, V_{qR} (q=u,d)$ 
which are used to diagonalize  
the $6 \times 6$ mass matrices for the up-quark sector and down-quark sector.
These unitary matrices are related to the mixing matrices for the
charged currents and neutral currents so that 
the $3 \times 6 $ submatrix of the unitary matrices in $V_{qL},V_{qR}$
enters into both 
charged currents $\V_L, \V_R$ and the neutral currents
 $\Z_{uL},\Z_{uR},\Z_{dL},$ and $\Z_{dR}$. The CP violation of 
the tree-level FCNC is determined by the imaginary parts
of the triangular matrices. Therefore, we conclude that
the FCNC is determined by the WB invariants $I_1-I_8$.
The mixing matrices
for the charged currents also depend on the $3 \times 3$ unitary matrices
$U_L$ and $U_R$ defined by Eqs. ({\ref{eq:downL}}) and (\ref{eq:newbasis}).

We obtain the mixing matrix elements by carrying out the approximate diagonalization so that we have some insight on the mixings and CP violation in terms of the mass eigenstates. 
As discussed in Sec. VI, we identify all 19 CP violating 
phases for the three-generation model in the couplings of charged currents and the neutral currents in terms of the mass eigenstates.
\appendix

\section{Exact formulas of matrices for the diagonalization}
In this appendix, we give the derivation of the formulas for
Eq.(\ref{eq:relations}). We also collect the formulas
that the submatrices of $V_{qL}$ satisfy [Eqs.(\ref{eq:d1})-(\ref{eq:d10})].
The proof of the formulas is given below.
One starts with  Eqs.(\ref{eq:vulvur}) and
(\ref{eq:vdlvdr})
for the diagonalization of the mass matrix
${\cal M_Q}$, which leads to the following relation:
\bea
V_{qR}&=&{\cal M_Q^\dagger} V_{qL} \begin{pmatrix} \frac{1}{d_q} & 0 \\
                                              0 & \frac{1}{\tD_Q} \end{pmatrix}
\nn \\
&=& \begin{pmatrix}
y_{\Delta qR} v_R S_{qL}/d_q & y_{\Delta qR} v_R T_{QL}/\tD_Q \\
(y_{\Delta qL}^\dagger v_L K_{qL}  + D_{Q} S_{qL})/d_q & (y_{\Delta qL}^\dagger v_L R_{qL} + 
D_Q T_{QL})/\tD_Q \end{pmatrix}, 
\label{eq;16}
\eea
where $(q, Q, \Q)$ denotes $(u, U, \U)$ or $(d,D,\D)$.
Equation.(\ref{eq;16}) leads to the formulas in Eq.(\ref{eq:relations}).
Since $V_{qL}$ satisfies the eigenvalue equation,
\bea
V_{qL}^\dagger {\cal M_Q M_Q^\dagger} V_{qL} =\begin{pmatrix} d_q^2 & 0\\ 
                        0 & \tD_Q^2 \end{pmatrix},
\eea
where 
\bea
{\cal M_Q M_Q^\dagger}=\begin{pmatrix} y_{\Delta qL} y_{\Delta qL}^\dagger v_L^2 & y_{\Delta qL} v_L D_Q \\
                                         D_Q v_L y_{\Delta qL}^\dagger & y_{\Delta qR}^\dagger y_{\Delta qR} v_R^2+ D_Q^2
\end{pmatrix},
\eea
the submatrices of $K_{qL}, S_{qL}, R_{qL}$, and $T_{QL}$ satisfy
\bea
&&y_{\Delta qL} y_{\Delta qL}^\dagger v_L^2 K_{qL} + y_{\Delta qL} v_L D_Q S_{qL} = K_{qL} d_q^2,  \label{eq:d1}\\
&& D_Q y_{\Delta qL}^\dagger v_L K_{qL} +(y_{\Delta qR}^\dagger y_{\Delta qR} v_R^2+D_Q^2) S_{qL}=S_{qL} d_q^2, \label{eq:d2}  \\
&&y_{\Delta qL} y_{\Delta qL}^\dagger v_L^2 R_{qL} + y_{\Delta qL} v_L D_Q T_{QL} = R_{qL} \tD_Q^2,  \label{eq:d3}\\
&& D_Q y_{\Delta qL}^\dagger v_L R_{qL} +(y_{\Delta qR}^\dagger y_{\Delta qR} v_R^2+D_Q^2) T_{QL}=T_{QL} \tD_Q^2. \label{eq:d4}
\eea
They also satisfy the unitarity conditions.
$V_{qL}^\dagger V_{qL} =1$ leads to
\bea
&&K_{qL}^\dagger K_{qL}+ S_{qL}^\dagger S_{qL} =1, \label{eq:d5} \\
&&R_{qL}^\dagger R_{qL} + T_{QL}^\dagger T_{QL} =1, \label{eq:d6} \\ 
&& K_{qL}^\dagger R_{qL} + S_{qL}^\dagger T_{QL} =0. \label{eq:d7}
\eea
$V_{qL} V_{qL}^\dagger=1$ leads to  
\bea
&& K_{qL} K_{qL}^\dagger + R_{qL} R_{qL}^\dagger =1, \label{eq:d8} \\
&& S_{qL} S_{qL}^\dagger + T_{QL} T_{QL}^\dagger =1, \label{eq:d9}\nn \\
&& K_{qL} S_{qL}^\dagger + R_{qL} T_{QL}^\dagger=0. \label{eq:d10}
\eea
Using the equations above, one can rewrite $V_{qR}$ as
\bea
V_{qR}=\begin{pmatrix} y_{\Delta qR} v_R S_{qL}/d_q & y_{\Delta qR} v_R T_{QL}/\tD_Q \\
                   \frac{1}{y_{\Delta qL}v_L} K_{qL} d_q & \frac{1}{y_{\Delta qL} v_{qL}} R_{qL} \tD_Q \end{pmatrix}.
\label{eq:vqr}
\eea
\section{Derivation of the approximate formulae }
In this appendix, we show the derivation for the approximate
formulas Eqs.(\ref{eq:kul})-(\ref{eq:ardr}) for $V_{qL}$ and $V_{qR}$.
The approximate diagonalization of the mass matrix of the universal
seesaw model has been  carried out in the previous works [9,20]. Compared to the previous works,
we relax the condition imposed on the singlet mass parameter $M_T$.
In this work, we do not assume that the parameter is very small compared 
to the $SU(2)_R$ breaking scale. We also keep all the CP violating parameters
in the approximation so that we can keep track of the CP violating phases in the
mixing matrices.

We show the derivation for the uptype quark case. The derivation for
the down-quark sector follows in the same way as that of the up-quark sector.
The submatrices of $K_{uL}, S_{uL}, R_{uL}$, and $T_{uL}$ satisfy
Eqs.(\ref{eq:d1})-(\ref{eq:d4}).
One also notes that $S_{uL}$ and $R_{uL}$ are smaller than $K_{uL}$ and $T_{uL}$.
Let us start with the Hermitian matrix,
\bea
H_\U=y_{\Delta uR}^\dagger y_{\Delta uR} v_R^2+D_U^2.
\eea
By neglecting the small contribution
proportional to $R_{uL}$, Eq.(\ref{eq:d4}) is rewritten as
\bea
T_{UL}^{0 \dagger} H_\U T_{UL}^0=\tD_U^{2},
\eea
where we denote the leading form for $T_{uL}$ as $T_{uL}^0$
and we use $T_{uL}^{0 \dagger} T_{uL}^0=1$.
The dominant term of $H_\U$ is
\bea
&&H_\U \sim D_{0U}^2 \equiv \begin{pmatrix} M_U^2 & 0 & \\
                            0 & M_C^2 & 0 \\
                            0 & 0 & D_{T}^2 \end{pmatrix},\label{eq:d0u}\\
&&D_{T}=\sqrt{y_{uR3}^2 v_R^2+M_T^2} \label{eq:d033}.
\eea
Therefore,
\bea
T_{uL}^0 \simeq 1, \quad \tD_U \simeq D_{0U}.
\label{eq:TuL}
\eea
Then one can solve Eq.(\ref{eq:d3}) for $R_{uL}$,
\bea
R_{uL} & \simeq &y_{\Delta uL} v_L \frac{D_{U}}{D_{0U}^2}.
\label{eq:arul}
\eea
In Eq.(\ref{eq:d2}), by neglecting the term
proportional to $d_{uL}^2$, one can solve $S_{uL}$ as
\bea
S_{uL} \simeq -\frac{1}{H_\U}D_U y_{\Delta uL}^\dagger v_{L} K_{uL}.
\label{eq:asul}
\eea
Then, $V_{uL}$ is approximately given as
\bea
V_{uL}=\begin{pmatrix} K_{uL} &  y_{\Delta uL} v_L \frac{D_U}{D_{0U}^2} \\
                    -\frac{D_Uv_L}{D_{0U}^2} y_{\Delta uL}^\dagger K_{uL} & 1 
\end{pmatrix},
\label{eq:avul}
\eea
where we use the approximation ${\cal H}_\U \simeq D_{0U}^2$.
One can also substitute $S_{uL}$ in 
Eq.(\ref{eq:asul}) and $R_{uL}$ in
Eq.(\ref{eq:arul}) into Eq.(\ref{eq:vqr}) and obtain
$V_{uR}$,
\bea
V_{uR}=\begin{pmatrix} -y_{\Delta uR} v_R\frac{D_U}{D_{0U}^2} 
y_{\Delta uL}^\dagger 
K_{uL}\frac{v_L}{d_u} & y_{\Delta uR} \frac{v_R}{D_{0U}}\\
\frac{1}{y_{\Delta uL} v_L} K_{uL} d_u & \frac{D_U}{D_{0U}} \end{pmatrix}.
\label{eq:avur}
\eea
Both $V_{uL}$ and $V_{uR}$ can be determined once the submatrix
$K_{uL}$ is fixed. The equation which determines $K_{uL}$ is obtained 
as
\bea
v_L^2 \mathscr{H} K_{uL} =K_{uL} d_u^2,
\label{eq:KuL}
\eea
where $\mathscr{H}$ is defined as
\bea
\mathscr{H}=y_{\Delta uL}(1-D_U \frac{1}{H_\U} D_U) y_{\Delta uL}^\dagger.
\eea
When deriving Eq.(\ref{eq:KuL}), Eq.(\ref{eq:asul}) is substituted into 
Eq.(\ref{eq:d1}).
Using the approximation
$ \det H_\U \simeq  H_{\U 11} H_{\U 22} H_{\U 33} 
\simeq M_U^2 M_C^2 H_{\U 33}, \quad
H_{\U 11} \simeq M_U^2, \quad H_{\U 22} \simeq M_C^2$,
and
by keeping the leading term in each matrix element, one obtains
\bea
&&\mathscr{H}/v_R^2=\nn \\
&&\begin{pmatrix} \frac{y_{uL1}^2}{M_U^2} (y_{uR1}^2+|y_{uR21}|^2
+\frac{|M_T|^2 |y_{uR31}|^2}{D_{T}^2})
& \frac{y_{uL1} y_{uL2}}{M_U M_C}(y_{uR21}^\ast y_{uR2}+
\frac{y_{uR31}^\ast y_{uR32} M_T^2}{D_{T}^2})&  
\frac{M_T}{M_U} \frac{y_{uL1} y_{uL3} y_{uR31}^\ast y_{uR3}}{D_{T}^2}\\ 
\frac{y_{uL1} y_{uL2}}{M_U M_C}(y_{uR21} y_{uR2}+
\frac{y_{uR32}^\ast y_{uR31} M_T^2}{D_{T}^2})
 &\frac{y_{uL2}^2}{M_C^2}(y_{uR2}^2 +\frac{|y_{uR32}|^2 M_T^2}
{D_{T}^2})& \frac{M_T}{M_C} 
\frac{y_{uL2}y_{uL3}y_{uR32}^\ast y_{uR3}}{D_{T}^2}\\
\frac{M_T}{M_U} 
\frac{y_{uL1} y_{uL3} y_{uR3}y_{uR31}}{D_{T}^2} & \frac{M_T}{M_C}
\frac{y_{uL2}y_{uL3} y_{uR3}y_{uR32}}{D_{T}^2}
& \frac{y_{uL3}^2 y_{uR3}^2}{D_{T}^2}
\end{pmatrix}. \nn \\
\eea
Now we solve the eigenvalue equation for $K_{uL}$. 
The eigenvalues of $\mathscr{H}$ are related 
to the up, charm, and top quark masses squared.
We write the eigenvalue equation as
\bea
\mathscr{H}
\left(
\begin{array}{cc}
\mathbf{u}\\
u_3
\end{array}
\right)
=\frac{m_i^2}{v_L^2}
\left(
\begin{array}{cc}
\mathbf{u}\\
u_3
\end{array}
\right),
\label{eq:eigen}
\eea
where $\mathbf{u}^T=(u_1,u_2)$ and $i=u,c,t$. We can rewrite  
Eq.(\ref{eq:eigen}) as
\bea
\left(
\begin{array}{cc}
\H_{11}& \H_{12} \\
\H_{12}^* & \H_{22}
\end{array}
\right)
\mathbf{u}+
\left(
\begin{array}{cc}
\H_{13}\\
\H_{23}
\end{array}
\right)
u_3
=\frac{m_i^2}{v_L^2} {\bf u},
\label{eq:uu3}
\eea
\bea
\left(
\begin{array}{cc}
\H_{13}^*& \H_{23}^* 
\end{array}
\right)
\cdot \mathbf{u}+\H_{33} u_3=\frac{m_i^2}{v_L^2} u_3.
\label{eq:u3u}
\eea
We first determine the eigenvalue and eigenvector for the top quark.
Because $\frac{m_t^2}{v_L^2} \gg \mH_{ij} (i,j=1,2)$, 
one can solve Eq.(\ref{eq:uu3}),
\bea
\mathbf{u}=\frac{v_L^2}{m_t^2}\begin{pmatrix}
\H_{13}\\
\H_{23}
\end{pmatrix} u_3.
\eea
Since $\mathbf{u}\ll u_3$, the top quark mass is approximately given as
\bea
m_t = v_L \sqrt{\H_{33}}= y_{uL3} v_L \frac{y_{uR3} v_R}{D_{T}}.
\eea 
The corresponding eigenvector for the top quark is given as
\bea
{\bf v}_t=\frac{1}{\sqrt{1+|\frac{\H_{13}}{\H_{33}}|^2+|\frac{\H_{23}}{\H_{33}}|^2}}
\begin{pmatrix}
\frac{\H_{13}}{\H_{33}}\\
\frac{\H_{23}}{\H_{33}}\\
1
\end{pmatrix}.
\label{eq42}
\eea
We ignore the correction in the following analysis since
the normalization factor of Eq.(\ref{eq42}) is close to $1$.
The other two eigenvectors ${\bf v}_u$ and ${\bf v}_c$ correspond to
the eigenvalues $m_u^2/v_L^2$ and $m_c^2/v_L^2$.
For the small eigenvalues, Eq.(\ref{eq:u3u}) can be solved as
\bea
u_3=-\frac{\begin{pmatrix}
\H_{13}^* & \H_{23}^* \end{pmatrix}
\cdot \mathbf{u}}{\H_{33}}.
\eea
Substituting the relation into Eq.(\ref{eq:uu3}), 
one obtains the following equation for the up and
charm quarks:
\bea
\begin{pmatrix} h_{11} & h_{12} \\
                h_{12}^\ast & h_{22} \end{pmatrix}
\mathbf{u} =\frac{m_i^2}{v_L^2} \mathbf{u},
\eea
where $i=u,c$ and $h_{ij}$ $(i,j=1,2)$ is defined as
\bea
h_{ij}=\H_{ij}-\H_{i3}\frac{1}{\H_{33}}\H_{3j}.
\eea
The components of $h$ are written explicitly, 
\bea
h_{11}&=&\frac{y_{uL1}^2 v_R^2 (y_{uR1}^2+|y_{uR21}|^2)}{M_U^2}, \nn \\
h_{12}&=&\frac{y_{uL1} y_{uL2} v_R^2(y_{uR21}^\ast y_{uR2})}{M_U M_C} 
,\nn \\
h_{22}&=&\frac{y_{uL2}^2 v_R^2 y_{uR2}^2}{M_C^2}.
\eea
Then for the up quark, the eigenvalue and the eigenvector are given as
\bea
&&
m_u=y_{uL1}y_{uR1} \frac{v_L v_R}{M_U},\quad
\mathbf{v}_u=
\begin{pmatrix}
1\\
-\frac{h_{12}^*}{h_{22}}\\
-\frac{\H_{13}^*-\frac{h_{12}^*}{h_{22}}\H_{23}^*}{\H_{33}}
\end{pmatrix}, 
\eea
and for charm quark, they are given as
\bea
&& m_c=y_{uL2}y_{uR2} \frac{v_L v_R}{M_C},\quad \mathbf{v}_c=\begin{pmatrix}
\frac{h_{12}}{h_{22}} \\
1 \\
-\frac{\H_{13}^\ast \frac{h_{12}}{h_{22}}+ \H_{23}^\ast}{\H_{33}}
\end{pmatrix} \simeq
\begin{pmatrix}
\frac{h_{12}}{h_{22}} \\
1 \\
-\frac{\H_{23}^\ast}{\H_{33}}
\end{pmatrix}.
\eea
$K_{uL}$ is written in terms of the eigenvectors,
\bea
K_{uL}&=&\begin{pmatrix} \bf{v}_{u} & \bf{v}_{c} & \bf{v}_{t}.\end{pmatrix}.
\label{eq:akul}
\eea
Similarly, the downtype mixing matrices
$K_{dL}$ and $R_{dL}$ are obtained.
The eigenvalues for the quark masses agree with the ones obtained in
Ref. \cite{Kiyo:1998zm}.
\section{parametrization of the Yukawa matrix in terms of a product of the
unitary matrix and triangular matrix}
In this appendix, we give proof of the parametrization
of the general $ 3 \times 3$ Yukawa matrices in terms of the product
of the 
unitary matrices and triangular matrices. The decomposition
and the parametrization are used in Eqs.(\ref{eq:product}) 
and (\ref{eq:productR}).
The general $3\times 3$ complex matrix of the Yukawa coupling
$Y$ with $\det Y \ne 0$ is written in terms of three independent complex vectors in $C^3$
${\bf y^0}_i$, $(i=1-3)$ 
as follows:
\bea
Y&=&\begin{pmatrix} {\bf y^0_1}& {\bf y^0_2}& {\bf y^0_3} \end{pmatrix} \nn \\ 
 &=&P(\alpha_1, \alpha_2, \alpha_3) \begin{pmatrix} {\bf y_1}& {\bf y_2}& 
{\bf y_3} \end{pmatrix}, \label{eq:decom1} \\
\alpha_i&=&\arg({\bf y^0_3}_i), 
\eea
where
$P(\alpha_1,\alpha_2,\alpha_3)$ is a diagonal 
unitary matrix defined in Eq.(\ref{eq:P}),
and ${\bf y_3}$ is a real vector in $R^3$. 
Then, we show $Y$ can be parametrized as
\bea
Y&=&P(\alpha_1,\alpha_2,\alpha_3) 
V(\theta_1,\theta_2,\theta_3, \delta)P(\alpha,\beta,0) Y_{\Delta}, 
\label{eq:param}
\eea
where $V(\theta_1, \theta_2, \theta_3, \delta)$ is a Kobayashi-Maskawa-type
parametrization for the unitary matrix and $Y_{\Delta}$ is a lower 
triangular matrix
with real diagonal elements,
{\small
\bea
&&V(\theta_1,\theta_2,\theta_3, \delta)=\nn \\
&&\begin{pmatrix} 
\cos \theta_3 \cos \theta_2 \cos \theta_1
 +\sin \theta_2 \sin \theta_1 e^{i \delta}
& \cos \theta_3 \cos \theta_2 \sin \theta_1
 -\sin \theta_2 \cos \theta_1 e^{i \delta} & \sin \theta_3 \cos \theta_2 \\
\cos \theta_3 \sin \theta_2 \cos \theta_1 -\cos \theta_2 \sin \theta_1 e^{i \delta}
&\cos \theta_3 \sin \theta_2 \sin \theta_1 + \cos \theta_2 \cos \theta_1 e^{i \delta}
& \sin \theta_3 \sin \theta_2
\\ 
-\sin \theta_3 \cos \theta_1 &
-\sin \theta_3 \sin \theta_1
&
 \cos \theta_3
\end{pmatrix},
\label{eq:ckm}\\
&& Y_{\Delta}= 
\begin{pmatrix} y_{\Delta11} & 0 & 0 \\
                              y_{\Delta 21}& y_{\Delta 22} & 0 \\
                              y_{\Delta 31}& y_{\Delta 32} & y_{\Delta 33}
\end{pmatrix}=
\begin{pmatrix}\cos \theta_{21} |{\bf y_1}|  & 0 & 0 \\
\sin \theta_{21} \cos \theta_{31} e^{i \phi_{21}} |{\bf y_1}|& 
\cos \theta_{32} |{\bf y_2}| & 0 \\ 
\sin \theta_{21} \sin \theta_{31} e^{i \phi_{31}} |{\bf y_1}|& 
\sin \theta_{32} e^{i \phi_{32}} |{\bf y_2}| & |{\bf y_3}| \end{pmatrix}
\label{eq:tria}.
\eea}
Equation.(\ref{eq:param}) shows a well-known result \cite{Morozumi:1997af};
i.e., the matrix $Y$ is written as the product of the unitary
matrix and the triangular matrix.
Here we show that a particular form of the parametrization
including some phases, angles, etc., shown in Eq.(\ref{eq:param}) is 
indeed a generic parametrization.
In this parametrization, there are nine real parts constructed by six angles,
\bea
\theta_1, \theta_2, \theta_3, \theta_{21}, \theta_{32}, \theta_{31}, 
\eea
and three norms of the complex vectors $|\bf y^0_i|=|\bf y_i|$, $(i=1,2,3)$.
The nine phases are given by
\bea
\alpha_1, \alpha_2, \alpha_3, \alpha,\beta, \delta,
\phi_{21}, \phi_{32}, \phi_{31}. 
\eea
Now we prove the parametrization is completely general.
One can start with
\bea
P(-\alpha_1,-\alpha_2,-\alpha_3)Y=\begin{pmatrix} {\bf y_1} & {\bf y_2} & {\bf y_3} \end{pmatrix},
\label{eq:PY}
\eea 
where ${\bf y_3}$ is a real vector in $R^3$. Further one can take out the norm
of ${\bf y_i}$ as
\bea
\begin{pmatrix} {\bf y_1} & {\bf y_2} & {\bf y_3} \end{pmatrix} 
 =\begin{pmatrix} {\bf v_1} & {\bf v_2} & {\bf v_3} \end{pmatrix}
\begin{pmatrix} |{\bf y_1}| & 0 & \\ 0 & |{\bf y_2}| & 0 \\
0 & 0 & |{\bf y_3}| \end{pmatrix}.
\label{eq:yv}
\eea
Note that ${\bf v}_i$ $(i=1-3)$ 
are normalized 
as ${\bf v_i^\dagger \cdot  v_i}=1$ but are not necessarily orthogonal.
${\bf v_3}$ is a real normalized vector, which implies
\bea
{\bf v_3}={\bf e_3}=\begin{pmatrix} \sin \theta_3 \cos \theta_2 \\
                        \sin \theta_3 \sin \theta_2 \\
                         \cos \theta_3 \end{pmatrix}.
\eea
We first show the general parametrization for orthonormal basis vectors
${\bf e_1}, {\bf e_2},$ which are orthogonal
to ${\bf e_3}$ in complex $C^3$ satisfying ${\bf e_i^\dagger \cdot e_j}=\delta_{ij}$.
Since ${\bf e_i} (i=1,2)$ are orthogonal to ${\bf e_3}$, 
both real parts and imaginary parts of ${\bf e_i}$ ($i=1,2$) are orthogonal to
${\bf e_3}$. 
Therefore, they are unitary superpositions of the two real orthogonal
vectors ${\bf e^0_1}$ and ${\bf e^0_2}$,
\bea
&& \begin{pmatrix} {\bf e_1} & {\bf e_2} \end{pmatrix}
=\begin{pmatrix} {\bf e^0_1} & {\bf e^0_2} \end{pmatrix} 
U,\nn \\
&&{\bf e^0_1}=\begin{pmatrix} \cos \theta_3 \cos \theta_2 \\
\cos \theta_3 \sin \theta_2 \\ -\sin \theta_3 \end{pmatrix}
, \quad {\bf e^0_2}=\begin{pmatrix} -\sin \theta_2 \\
\cos \theta_2 \\ 0 \end{pmatrix},
\eea
where two-by-two unitary matrix denoted by $U$ can be parametrized as
\bea
U&\equiv& 
\begin{pmatrix}
1 & 0 \\
0& e^{i \delta} \end{pmatrix}\begin{pmatrix}
\cos \theta_1 & \sin \theta_1 \\
-\sin \theta_1& \cos \theta_1\end{pmatrix}\begin{pmatrix}
e^{i \alpha} & 0 \\
0& e^{i \beta} \end{pmatrix}
\eea
Then, one can write ${\bf e_2}$ and ${\bf e_1}$ as
\bea
{\bf e_2}&=& \begin{pmatrix} \cos \theta_3 \cos \theta_2 \sin \theta_1
 -\sin \theta_2 \cos \theta_1 e^{i \delta}
 \\
\cos \theta_3 \sin \theta_2 \sin \theta_1 + \cos \theta_2 \cos \theta_1 e^{i \delta}
\\ -\sin \theta_3 \sin \theta_1 \end{pmatrix} e^{i \beta}
,\nn \\
{\bf e_1}&=& \begin{pmatrix} \cos \theta_3 \cos \theta_2 \cos \theta_1
 +\sin \theta_2 \sin \theta_1 e^{i \delta}
 \\
\cos \theta_3 \sin \theta_2 \cos \theta_1 -\cos \theta_2 \sin \theta_1 e^{i \delta}
\\ -\sin \theta_3 \cos \theta_1 \end{pmatrix} e^{i \alpha}.
\eea
From ${\bf v_2}$ one can form the vector that is 
orthogonal to ${\bf e_3}$.
This vector can be identified with ${\bf e_2}$,
\bea
\frac{\bf v_2- e_3^T\cdot v_2 e_3}{\sqrt{1-|{\bf e_3^T \cdot v_2}|^2}}={\bf e_2}. 
\eea
Therefore, one can write ${\bf v_2}$ with the superposition,
\bea
{\bf v_2}&=& {\bf e_3^T \cdot v_2} {\bf e_3}+ \sqrt{1-|{\bf e_3^T \cdot v_2}|^2} {\bf e_2},
\nn \\
&=& \sin \theta_{32} e^{i \phi_{32}} {\bf e_3}+ \cos \theta_{32} {\bf e_2},
\label{eq:v2}
\eea
where we set $ {\bf e_3^T \cdot v_2}=\sin \theta_{32} e^{i \phi_{32}}$.
Next from ${\bf v_1}$, one can form the vector that is orthogonal to
${\bf e_3}$ and ${\bf e_2}$. This can be identified as ${\bf e_1}$,
\bea
&&\frac{\bf v_1-e_3^T\cdot v_1 e_3-e_2^\dagger \cdot v_1 e_2}{\sqrt{1-|{\bf e_3^T \cdot v_1}|^2-
|\bf e_2^\dagger \cdot v_1}|^2}=\bf{e_1}, \nn \\
&&{\bf v_1}=\cos \theta_{21} {\bf e_1} +
\sin \theta_{21} \cos \theta_{31} e^{i \phi_{21}} {\bf e_2} 
+\sin \theta_{21} \sin \theta_{31} e^{i \phi_{31}} {\bf e_3}, 
\label{eq:v1}
\eea
where one sets 
${\bf e_3^T\cdot v_1}=\sin \theta_{21} \sin \theta_{31} e^{i \phi_{31}}$
and ${\bf e_2^\dagger \cdot v_1}=\sin \theta_{21} \cos \theta_{31} e^{i \phi_{21}}$.
We summarize the relation $({\bf e_1,e_2,e_3})$ with $({\bf v_1,v_2, v_3})$
using Eqs.(\ref{eq:v2}) and (\ref{eq:v1}).
\bea
\begin{pmatrix} {\bf v_1} & {\bf v_2} & {\bf v_3} \end{pmatrix}
=\begin{pmatrix} {\bf e_1} & {\bf e_2} & {\bf e_3} \end{pmatrix}
\begin{pmatrix}\cos \theta_{21}  & 0 & 0 \\
\sin \theta_{21} \cos \theta_{31} e^{i \phi_{21}} & \cos \theta_{32}  & 0 \\ 
\sin \theta_{21} \sin \theta_{31} e^{i \phi_{31}} & \sin \theta_{32} e^{i \phi_{32}} & 1 \end{pmatrix}.
\label{eq:ve}
\eea
Note that the unitary matrix $({\bf e_1, e_2, e_3})$ is written in terms of 
three angles and three phases as
\bea
({\bf e_1, e_2, e_3})
= V(\theta_1, \theta_2,\theta_3, \delta) 
P(\alpha, \beta,0).
\label{eq:e1e2e3}
\eea
We substitute the relation Eq.(\ref{eq:e1e2e3}) into Eq.(\ref{eq:ve}). 
Then one obtains
{\small
\bea
\begin{pmatrix} {\bf v_1} & {\bf v_2} & {\bf v_3} \end{pmatrix}
=
V(\theta_1, \theta_2,\theta_3, \delta) 
P(\alpha, \beta,0)
\begin{pmatrix}\cos \theta_{21}  & 0 & 0 \\
\sin \theta_{21} \cos \theta_{31} e^{i \phi_{21}} & \cos \theta_{32}  & 0 \\ 
\sin \theta_{21} \sin \theta_{31} e^{i \phi_{31}} & \sin \theta_{32} e^{i \phi_{32}} & 1 \end{pmatrix},
\eea}
which implies
\bea
P(-\alpha_1, -\alpha_2, -\alpha_3)Y 
=V(\theta_1, \theta_2, \theta_3, \delta) P(\alpha, \beta, 0)
Y_{\Delta}.
\label{eq:py}
\eea
One can easily derive Eq.(\ref{eq:param}) from Eq.(\ref{eq:py}). 
\begin{acknowledgments}
T. M. was supported by KAKENHI, Grant-in-Aid for 
Scientific Research(C) Grant No. 22540283 from JSPS, Japan.
\end{acknowledgments}

\end{document}